\documentclass[twocolumn,showpacs,aps,prd,superscriptaddress,10pt]{revtex4-2}

\usepackage[retainorgcmds]{IEEEtrantools}

\usepackage{ifpdf}
\usepackage[utf8]{inputenc}
\usepackage[UKenglish]{babel}
\usepackage{graphicx}
\usepackage{amsmath, amsthm, bm, amssymb}
\usepackage{mathrsfs}
\usepackage{multirow}
\usepackage{url}
\usepackage{tikz}
\usepackage{xspace}
\usetikzlibrary{positioning,calc}
\usepackage{subcaption}
\usepackage{color}
\usepackage{tabularx}
\usepackage{hhline}
\usepackage{mathtools}
\usepackage[colorlinks,allcolors=magenta]{hyperref}
\usepackage{siunitx}
\usepackage[commentmarkup=uwave]{changes}

\setdeletedmarkup{\color{red}\sout{#1}}

\def\updowntensor#1#2#3{#1^{#2}_{\hphantom{#2}#3}}
\def\updowndowntensor#1#2#3#4{#1^{#2}_{\hphantom{#2}#3#4}}

\definechangesauthor[color=purple,name={Leo Werneck}]{LW}
\definechangesauthor[color=violet,name={Samuel Cupp}]{SC}

\def\eqref#1{Eq.\,(\ref{#1})\xspace}
\def\alteqref#1{Eq.\,\ref{#1}\xspace}
\def\figref#1{Fig.\,\ref{#1}\xspace}

\def\secref#1{Sec.\,\ref{#1}\xspace}

\newcommand{\code}[1]{\textsf{#1}}

\newcommand{\eos}{EoS\xspace}
\newcommand{\eoss}{EoSs\xspace}

\newcommand{\bhah}{\code{BlackHoles@Home}\xspace}
\newcommand{\grhayl}{\code{GRHayL}\xspace}
\newcommand{\groovy}{\code{GRoovy}\xspace}

\newcommand{\grhaylhd}{\code{GRHayLHD}\xspace}
\newcommand{\grhaylmhd}{\code{IllinoisGRMHD}\xspace}

\newcommand{\fuka}{\code{FUKA}\xspace}
\newcommand{\kadath}{\code{KADATH}\xspace}

\newcommand{\igm}{\code{IllinoisGRMHD}\xspace}

\newcommand{\etk}{\code{Einstein Toolkit}\xspace}
\newcommand{\baikal}{\code{Baikal}\xspace}

\newcommand{\nrpy}{\code{NRPy}\xspace}

\def\reviewed#1{%
\ifthenelse{\equal{#1}{}}{\relax}{\noindent{\color{teal}\textbf{Reviewed: #1}}\par}}
\def\pending#1{%
\ifthenelse{\equal{#1}{}}{\relax}{\noindent{\color{red}\textbf{Pending: #1}}\par}}

\newcommand{\mathvar}[1]{\ensuremath{#1}}

\newcommand{\primv}{\mathvar{\bm{P}}\xspace}
\newcommand{\consv}{\mathvar{\bm{C}}\xspace}
\newcommand{\fluxv}{\mathvar{\bm{F}}\xspace}
\newcommand{\sourcev}{\mathvar{\bm{S}}\xspace}

\newcommand{\ye}{\mathvar{Y_\mathrm{e}}\xspace}
\newcommand{\yet}{\mathvar{\tilde{Y}_\mathrm{e}}\xspace}
\newcommand{\sqrtgamma}{\mathvar{\sqrt{\gamma}}\xspace}

\newcommand{\nb}{\mathvar{n_\mathrm{b}}\xspace}
\renewcommand{\ne}{\mathvar{n_\mathrm{e}}\xspace}
\newcommand{\mb}{\mathvar{m_\mathrm{b}}\xspace}

\newcommand{\RR}{\mathcal{R}}
\newcommand{\QQ}{\mathcal{Q}}

\newcommand{\gratio}{\xi}

\newcommand{\orcid}[1]{\href{https://orcid.org/#1}{\includegraphics[height=\fontcharht\font`\B]{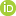}}}

\newcommand{\uidaho}{Department of Physics, University of Idaho, Moscow, ID 83843, USA}
\newcommand{\wvu}{Department of Physics and Astronomy, West Virginia University, Morgantown, WV 26506, USA}
\newcommand{\marshall}{Department of Mathematics and Physics, Marshall University, Huntington, WV 25755, USA}
\newcommand{\cgwc}{Center for Gravitational Waves and Cosmology, West Virginia University, Chestnut Ridge Research Building, Morgantown, WV 26505}

\begin{document}

\sloppy

\title{General relativistic hydrodynamics code for dynamical spacetimes with curvilinear coordinates, tabulated equations of state, and neutrino physics}

\author{Terrence~Pierre~Jacques~\orcid{0000-0002-8993-0567}}
\email{tp0052@mix.wvu.edu}
\affiliation{\wvu}
\affiliation{\cgwc}
\affiliation{\uidaho}

\author{Samuel~Cupp~\orcid{0000-0003-1758-8376}}
\email{scupp1@my.apsu.edu}
\affiliation{\uidaho}

\author{Leonardo~R.~Werneck~\orcid{0000-0002-4541-8553}}
\email{leonardo@uidaho.edu}
\affiliation{\uidaho}

\author{Samuel~D.~Tootle~\orcid{0000-0001-9781-0496}}
\email{sdtootle@uidaho.edu}
\affiliation{\uidaho}

\author{Maria~C.~Babiuc~Hamilton~\orcid{0000-0003-1878-2445}}
\email{babiuc@marshall.edu}
\affiliation{\marshall}
\affiliation{\cgwc}

\author{Zachariah~B.~Etienne~\orcid{0000-0002-6838-9185}}
\email{zetienne@uidaho.edu}
\affiliation{\uidaho}
\affiliation{\wvu}
\affiliation{\cgwc}

\begin{abstract}

Many astrophysical systems of interest to numerical relativity---such as rapidly rotating stars, black hole accretion disks, and core-collapse supernovae---exhibit near-symmetries. These systems generally consist of 
a strongly gravitating central object surrounded by an accretion disk, debris, and ejecta. Simulations can efficiently exploit the near-axisymmetry of these systems by reducing the number of points in the angular direction around the near-symmetry axis, enabling efficient simulations over seconds-long timescales with minimal computational expense. In this paper, we introduce \groovy, a novel code capable of modeling astrophysical systems containing compact objects by solving the equations of general relativistic hydrodynamics (GRHD) in full general relativity using singular curvilinear (spherical-like and cylindrical-like) and Cartesian coordinates. We demonstrate the code's robustness through a battery of challenging GRHD tests, ranging from flat, static spacetimes to curved, dynamical spacetimes. These tests further showcase the code's capabilities in modeling systems with realistic, finite-temperature equations of state and neutrino cooling via a leakage scheme. \groovy extensively leverages \grhayl, an open-source, modular, and infrastructure-agnostic general relativistic magnetohydrodynamics library built from the highly robust algorithms of \igm.
Long-term simulations of binary neutron star and black hole–neutron star post-merger remnants will benefit greatly from using a future \texttt{Charm++}-parallelized version of \groovy to study phenomena such as remnant stability, gamma-ray bursts, and nucleosynthesis.

\end{abstract}

\maketitle

\section{Introduction}
\label{sec:introduction}

The gravitational wave (GW) event GW170817~\cite{LIGO_GW170817, LIGO_2017ApJ} marked the first simultaneous direct detection of GWs and electromagnetic (EM) radiation from a binary neutron star (BNS) merger~\cite{LIGO_2017ApJ_EM_GWs}, launching a new era of multi-messenger astrophysics. This event, observed across X-ray, radio, and optical bands~\cite{Troja_2017, Troja_2018, Ruan_2018, Lamb_2018, Lamb_2019}, provided an unparalleled opportunity to test and refine theoretical models for the EM counterparts to BNS mergers~\cite{Metzger_2010, Metzger_2019_LRR, Sarin_2021, Ruiz_2016, Moesta_2020, Ciolfi_2020_outflows, Sun_2022, Kiuchi_2023}. Future observations are expected to further constrain the neutron star equation of state~\cite{Takami_2014, Radice_2018}, refine models of short gamma-ray burst (sGRB) emission, and deepen our understanding of neutrino physics and r-process nucleosynthesis~\cite{Cowan_2021, Siegel_2017_rprocess, Combi_2023_rprocess}.

While GW170817 remains the sole GW+EM event, the frequency of BNS and black hole-neutron star (BHNS) mergers, as inferred from current event rates observed by the LIGO-VIRGO-KAGRA collaboration~\cite{LIGO_GW190425, LIGO_2021ApJ_NSBH, LIGO_GWTC3}, underscores the need for compact object merger simulation catalogs. Given the non-linear nature of Einstein's field equations coupled to the equations of general relativistic magnetohydrodynamics (GRMHD), these catalogs must be built using self-consistent numerical relativity models that span both observational and theoretical uncertainties.

After the inspiral and merger phases, BNS and BHNS systems become nearly axisymmetric, with Keplerian fluid flows largely following grid lines in spherical or cylindrical coordinates, centered on the compact remnant at $r = 0$. When these flows are modeled on Cartesian numerical grids, they cross grid lines obliquely, resulting in spurious angular momentum loss and hindering the accuracy of post-merger simulations at low-to-moderate resolutions. Spherical or cylindrical coordinate systems enable more reliable post-merger simulations over timescales of seconds at lower resolutions. To date, several codes have been developed that adopt spherical coordinates~\cite{HARM, Noble_2006, Noble_2009, HARM3D, Liska_2022, Porth_2017, Cheong_2021, Mewes_2018, Mewes_2020, Ng_2023a, Stone_2008, Stone_2020, DelZanna_2007} for more accurate modeling of GRMHD and GRHD flows near compact objects.

Recently, ``hand-off'' scenarios have become popular, in which the inspiral and merger are modeled using codes with more general-purpose grids and Cartesian adaptive mesh refinement. Once the post-merger spacetime becomes largely stationary, the metric and GRMHD data are interpolated into a code that adopts spherical coordinates~\cite{Armengol_2022, Gottlieb_2023ApJ_a, Gottlieb_2023ApJ_b}. These specialized codes typically focus on static-spacetime black hole accretion disk simulations. Further work is still needed, however, as future studies will need to self-consistently model finite temperature effects and neutrino cooling of post-merger remnants across a vast parameter space of BNS and BHNS systems. Incorporating these processes will be essential for studying magnetized winds from accretion disks~\cite{Ciolfi_2020, Combi_2023, Curtis_2024} and understanding polar fluid flows that may evolve into ultra-relativistic jets over secular timescales.

Compact object mergers are not the only strongly gravitating astrophysical systems that are largely spherically or axially symmetric. Core-collapse supernova studies (e.g.,~\cite{O'Connor_2015, O'Connor_2018, Couch_2014, Schneider_2019, Moesta_2018, Morozova_2018, Burrows_2019}) have been conducted using spherical coordinates, either in 1D, 2D, or full 3D. Similarly, studies of direct collapse black holes~\cite{Chon_2018, Latif_2021, Luo_2018, Smidt_2018}, which couple hydrodynamics with a rapidly evolving gravitational field, may also benefit from using curvilinear coordinates. We also acknowledge the vast amount of research in modeling black hole accretion disks (e.g.,~\cite{Mocibrodzka_2009, Mocibrodzka_2014, Noble_2009, Shcherbakov_2012, Narayan_2012}) that often adopt spherical geometries to model accretion flows and jets.

To accurately and efficiently model this wide range of strongly gravitating and highly relativistic astrophysical systems that exhibit some degree of symmetry, we have developed \groovy, a new code within the \bhah/\nrpy framework~\cite{Ruchlin_2018, Etienne_nrpytutorial}. \groovy implements the equations of general relativistic hydrodynamics (GRHD) in full general relativity and supports spherical, cylindrical, and Cartesian coordinate systems.

Our code leverages the newly developed, open-source General Relativistic Hydrodynamics Library (\grhayl)~\cite{grhayl_github, Cupp_2024}, a highly modular redesign of the \igm code~\cite{Duez_IGM, Etienne_IGM, Werneck_IGM}. Through its straightforward, infrastructure-agnostic interface, \grhayl provides all the GRMHD algorithms of \igm, with each algorithm validated to roundoff error against the established \igm code. This modularity allows users to easily integrate only the algorithms needed for specific applications.

\grhayl's flexibility has allowed us to integrate \groovy's curvilinear-coordinate GRHD implementation with \igm's well-tested, realistic finite-temperature equations of state (EOS) and neutrino-leakage algorithms~\cite{Werneck_IGM}, enabling a wide range of physics within the GRHD regime. This integration facilitates modeling of neutrino-irradiated disk winds~\cite{Martin_2015, Perego_2014, Musolino:2024xqy} and outflows~\cite{Foucart_2021, Radice_2023}, bulk transport~\cite{Chabanov_2023a, Chabanov_2023b}, thermal transport~\cite{Paschalidis_2012, Raithel_2021, Hammond_2021, Fields_2023}, and the post-merger lifetime and stability of remnant neutron stars~\cite{Ciolfi_2017, Ruiz_2018, Tootle_2021, Tootle:2022, Papenfort2022a}. The future goal of the \texttt{Charm++}-based version of \groovy is to efficiently simulate the evolution of these systems over time, independent of magnetic field evolution.

Here, we showcase the capabilities of \groovy by modeling various hydrodynamical flows in different coordinate systems. Building upon the reference metric formulation of the GRHD equations presented in~\cite{Montero_2014, Mewes_2020, Cheong_2021, Xie_2024}, we incorporate the conservation of lepton number and source terms to model finite temperature effects and neutrino emission via a leakage scheme. We then numerically integrate these equations forward in time using a finite volume method, conducting a series of standard yet challenging numerical tests. These tests include two- and three-dimensional shock tests in spherical coordinates, neutrino leakage tests, and demonstrations of \groovy's ability to accurately model both non-rotating and rotating isolated neutron stars. Our results indicate that \groovy is highly robust in modeling flows within flat spacetimes, strongly curved static spacetimes, and fully dynamical spacetimes.

The remainder of this paper is organized as follows. Section~\ref{beq} outlines the GRHD equations our codes approximately solves. Section~\ref{num_int} details our numerical techniques and algorithms. In Section~\ref{code_tests}, we describe our setups and present the results of various code tests. We conclude in Section~\ref{conclusion}.

\section{Basic Equations}
\label{beq}

In this section, we present the core equations used to model astrophysical systems with our new code. We use geometrized units throughout this paper, where $G=c=M_\odot=1$, with $G$ as the gravitational constant and $c$ as the speed of light. Latin indices indicate spatial components, while Greek indices represent spacetime components. We express Einstein's equations in $3+1$ form, starting with the standard line element~\cite{Arnowitt:1962hi}:
\begin{equation}
    ds^2 = \left(-\alpha^2+\beta^i\beta_i\right)dt^2
        + 2\beta_i\,dt\, dx^i
        + \gamma_{ij}\,dx^i\,dx^j\;,
\end{equation}
where $\alpha$ is the lapse function, $\beta^j$ is the shift vector, and \mbox{$\gamma_{ij}=g_{ij}$} is the spatial part of the spacetime metric $g_{\mu\nu}$. We adopt the following conformal decomposition of the spatial metric~\cite{Baumgarte_Shapiro_2010}:
\begin{equation}
    \gamma_{ij} = e^{4\phi}\bar{\gamma}_{ij},
\end{equation}
where $\bar{\gamma}_{ij}$ is the conformal metric, $\psi\equiv e^\phi$ is the conformal factor, and $\phi$ is the conformal exponent. To evolve the spacetime in curvilinear coordinates, we further decompose the conformal metric as
\begin{equation}
    \bar{\gamma}_{ij} = \hat{\gamma}_{ij} + \epsilon_{ij},
\end{equation}
where $\hat{\gamma}_{ij}$ is the flat space reference metric in the chosen coordinate system, and $\epsilon_{ij}$ represents deviations from flat space that are not necessarily small. For example, in spherical coordinates, the flat reference metric is given by \mbox{$\hat{\gamma}_{ij} = \text{diag} \left(1, r^2, r^2 \sin^2\theta \right)$}. To simplify future expressions, we also define the quantity
\begin{equation}
\xi = e^{6\phi} \sqrt{\frac{\bar{\gamma}}{\hat{\gamma}}}.
\end{equation}
Following the Lagrangian BSSN formulation of~\cite{Brown_2009}, we set $\bar{\gamma}$ to remain constant in time. For numerical convenience, we choose this constant to be $\hat{\gamma}$, and enforce $\bar{\gamma} = \hat{\gamma}$ at each time step. As described in Section IV.~D of~\cite{Ruchlin_2018}, we ensure this condition by applying the algebraic correction
\begin{equation}
\bar{\gamma}_{ij} \to \left(\frac{\hat{\gamma}}{\bar{\gamma}}\right)^{1/3} \bar{\gamma}_{ij},
\end{equation}
to the conformal metric components at the end of each RK substep in our numerical evolution. In the following equations we retain the factor $\xi$ to maintain consistency with previous literature.

\groovy solves for the spacetime using the BSSN formulation~\cite{Shibata_1995, Baumgarte_1998, Brown_2009} of Einstein's equations with a reference metric prescription~\cite{Baumgarte_2013, Mewes_2018, Ruchlin_2018} implemented within the \nrpy code-generation framework. The \nrpy auto-generated BSSN reference-metric code has been validated in spherical coordinates against the independent code of~\cite{Baumgarte_2013} (see~\cite{Ruchlin_2018}) and in Cartesian coordinates against \texttt{ML\_BSSN} in the \etk using the \baikal thorn.

To couple the fluid and spacetime dynamics, we must also solve the GRHD equations:
\begin{align}
  \nabla_\mu\left(\nb u^\mu\right) &= 0\;,\label{eq:baryon_number_conservation} \\ 
  \nabla_\mu\left(\ne u^\mu\right) &= \RR/\mb\;,\label{eq:lepton_number_conservation} \\ 
  \nabla_\mu T^{\mu\nu} &= \QQ u^\nu\;,\label{eq:enmom_conservation}
\end{align}
which correspond to conservation laws for the baryon number $\nb$, lepton number $\ne$, and the GRHD stress-energy tensor
\begin{align*}
T^{\mu\nu} = \rho h u^\mu u^\nu + P g^{\mu\nu},
\end{align*}
respectively. Here, $\rho = \mb \nb$ is the fluid rest mass density, $\mb$ is the baryon mass, $h = 1 + \epsilon + P/\rho$ is the enthalpy, $\epsilon$ is the specific internal energy, $u^\mu$ is the fluid four-velocity, and $P$ is the pressure.
In addition, $\RR$ and $\QQ$ represent the source term contributions associated with neutrino emission via a leakage scheme (see e.g.,~\cite{Werneck_IGM, Siegel:2017jug}).

Following the detailed derivations of~\cite{Montero_2014, Mewes_2020}, we recast the conservation laws in a form appropriate for numerical integration. To bring the equations closer to the formulation presented in~\cite{Montero_2014} for the case of pure GRHD, we rewrite the above equations as follows:
\begin{align}
    \nabla_\mu\left(\rho u^\mu\right) &= 0\;,\label{eq:baryon_number_conservation_new} \\ 
    \nabla_\mu\left(\rho \ye u^\mu\right) &= \RR\;,\label{eq:lepton_number_conservation_new}
\end{align}
where we have introduced the electron fraction \mbox{$\ye = \ne/\nb$}. In the following subsections, we show how to rewrite these conservation laws using a reference metric formulation.

\subsection{Conservation of Lepton and Baryon Number}
\label{sec:lepton_baryon}

Following~\cite{Montero_2014} and using the identity
\begin{align}
    \nabla_\mu V^\mu = \frac{1}{\sqrt{|g|}} \partial_\mu \left(\sqrt{|g|} V^\mu\right)\;\label{eq:vector_identity},
\end{align}
we rewrite \eqref{eq:lepton_number_conservation_new} as
\begin{equation}
\frac{1}{\sqrt{-g}} \left[ \partial_t \left(\sqrt{-g} \rho \ye u^0 \right) + \partial_i \left(\sqrt{-g} \rho \ye u^i \right) \right] = \RR.\label{eq:ye_evolution_partial}
\end{equation}
Now, using $\sqrt{-g} = \alpha \sqrt{\gamma} = \alpha e^{6\phi} \sqrt{\bar{\gamma}}$, we can rewrite the second term in the square brackets of \eqref{eq:ye_evolution_partial} as
\begin{align*}
\partial_i \left(\alpha e^{6\phi} \sqrt{\bar{\gamma}} \rho \ye u^i \right) &= \partial_i \left(\sqrt{\hat{\gamma}} \alpha \gratio \rho \ye u^i \right).
\end{align*}
Note that since \eqref{eq:vector_identity} is valid for any metric, this last equation can be rewritten in terms of the covariant derivative associated with the reference metric $\hat{\gamma}_{ij}$, i.e.,
\begin{align*}
\partial_i \left(\sqrt{\hat{\gamma}} \alpha \gratio \rho \ye u^i \right) = \sqrt{\hat{\gamma}} \hat{\nabla}_i \left( \alpha \gratio \rho \ye u^i \right).
\end{align*}
Replacing the spatial terms in \eqref{eq:ye_evolution_partial} with this result and dividing by $\sqrt{\hat{\gamma}}$ gives the equation
\begin{equation*}
\partial_t \left(\alpha \gratio \rho \ye u^0 \right) + \hat{\nabla}_i \left( \alpha \gratio \rho \ye u^i \right) = \alpha \gratio \RR\;.
\end{equation*}
We define the conserved baryonic density as \mbox{$D \equiv W\rho$}, where \mbox{$W = \alpha u^{0}$} is the Lorentz factor between observers in the fluid and normal frames, the conserved electron fraction as $D \ye$, and the flux as
\begin{equation*}
\left( f_{\yet} \right)^i \equiv \alpha \gratio \rho \ye u^i = \gratio D\ye v^i = \yet v^i. 
\end{equation*}
Here, \mbox{$v^i = {u^i} / {u^0}$} is the fluid three-velocity and \mbox{$\yet \equiv \gratio D\ye$} is the ``densitized'' conserved electron fraction. We thus obtain the evolution equation
\begin{equation}
\partial_t \yet + \hat{\nabla}_i \left( f_{\yet} \right)^i = \alpha \gratio \RR.
\label{eq:consv_ye}
\end{equation}
Note that Eqs.~(\ref{eq:baryon_number_conservation_new})~and~(\ref{eq:lepton_number_conservation_new}) have similar mathematical forms, with the main difference being that \eqref{eq:baryon_number_conservation_new} does not have a source term. Introducing the ``densitized'' conserved density \mbox{$\tilde{D} \equiv \gratio D$}, it immediately follows that
\begin{eqnarray}
\partial_t \tilde{D} + \hat{\nabla}_i \left( f_{\tilde{D}} \right)^i &=& 0 \\
\implies \partial_t \tilde{D} + \partial_i \left( \tilde{D}v^{i} \right) &=& -\hat{\Gamma}^i_{i j} \tilde{D}v^{j},
\label{eq:consv_density}
\end{eqnarray}
where $\hat{\Gamma}^i_{i j}$ is the connection associated with the reference metric, and
\begin{equation}
\left( f_{\tilde{D}} \right)^i \equiv \gratio D v^i = \tilde{D}v^{i}.
\end{equation}
In Cartesian coordinates, connections vanish and we recover the Valencia formalism, where mass is conserved to machine precision. In curvilinear coordinates, source terms appear and lead to mass being conserved to truncation error. The same conclusion holds for the electron fraction evolution equation (\alteqref{eq:consv_ye}). These truncation errors will converge away in the continuum limit of increased resolution. Thus, as long as our numerical scheme is convergent at the expected order, which we demonstrate in Section \ref{code_tests}, this consequence is not expected to affect our numerical solutions at sufficient resolution.

\subsection{Conservation of Momentum}

To derive the Euler equation from \eqref{eq:enmom_conservation}, we begin with the identity
\begin{align*}
\nabla_{\nu} \updowntensor{A}{\nu}{\mu} = \frac{1}{\sqrt{-g}} \partial_{\nu} \left(\sqrt{-g} \updowntensor{A}{\nu}{\mu} \right) - {^{(4)}}\Gamma^{\rho}_{\nu \mu} \updowntensor{A}{\nu}{\rho},
\end{align*}
valid for some tensorial quantity $\updowntensor{A}{\nu}{\mu}$, where ${^{(4)}}\Gamma^{\rho}_{\nu \mu}$ is the metric connection associated with $g_{\mu\nu}$. 
We then take a spatial projection of \eqref{eq:enmom_conservation}, i.e.,
\begin{eqnarray*}
g_{i\nu} \QQ u^{\nu}  &=& g_{i\nu} \nabla_{\mu} T^{\mu\nu} = \nabla_{\mu} \big( g_{i\nu} T^{\mu\nu} \big) \\
&=& \frac{1}{\sqrt{-g}} \partial_{\mu} \big(\sqrt{-g} T^{\mu}_{\hphantom{\mu}i} \big) -  {^{(4)}}\Gamma^{\mu}_{i\nu} T_{\hphantom{\nu}\mu}^{\nu} \\
&=& \frac{1}{\sqrt{-g}} \left[ \partial_t \big(\sqrt{-g} T^{0}_{\hphantom{0}i} \big) + \partial_j \big(\sqrt{-g} T^{j}_{\hphantom{j}i} \big) \right] \\
&& - {^{(4)}}\Gamma^{\mu}_{i\nu} T_{\hphantom{\nu}\mu}^{\nu} .
\end{eqnarray*}
Similar to the approach in \ref{sec:lepton_baryon}, we expand terms with spatial derivatives as follows:
\begin{align*}
\partial_j \Big(\sqrt{-g} \updowntensor{T}{j}{i} \Big) &= \partial_j \Big(\sqrt{\hat{\gamma}} \alpha \gratio \updowntensor{T}{j}{i} \Big) \\
&= \sqrt{\hat{\gamma}} \hat{\nabla}_j \Big( \alpha \gratio \updowntensor{T}{j}{i} \Big) + \alpha e^{6\phi} \sqrt{\bar{\gamma}} \hat{\Gamma}^k_{ij} \updowntensor{T}{j}{k}.
\end{align*}
Now, using this expansion in the previous equation and replacing $\sqrt{-g}$, then dividing by $\sqrt{\hat{\gamma}}$, we obtain
\begin{align*}
    \partial_t \Big(\alpha \gratio \updowntensor{T}{0}{i} \Big) &+ \hat{\nabla}_j \Big( \alpha \gratio \updowntensor{T}{j}{i} \Big) \\
    &= -\alpha \gratio \Big( \hat{\Gamma}^k_{ij} \updowntensor{T}{j}{k} - {^{(4)}}\Gamma^{\mu}_{i\nu} \updowntensor{T}{\nu}{\mu}  \Big) + \alpha \gratio \QQ u_{i}.
\end{align*}
Since the expansion of the source terms has been done in~\cite{Montero_2014}, we refer the reader there for more details. Defining the densitized momentum as $\tilde{S_i} \equiv \alpha \gratio \updowntensor{T}{0}{i}$, the flux term as $ \left( f_{\tilde{S}} \right)^j_i \equiv \alpha \gratio \updowntensor{T}{j}{i}$, and the source term as
\begin{align}
    L_i &\equiv -T^{00} \alpha \partial_i \alpha + \updowntensor{T}{0}{k} \hat{\nabla}_i \beta^k \\ \nonumber
    &+ \frac{1}{2} \hat{\nabla}_i \gamma_{jk} \left( T^{00} \beta^j \beta^k + 2 T^{0j} \beta^k +  T^{jk} \right),
\end{align}
we write the equations in compact form
\begin{align}
\partial_t \tilde{S}_i + \hat{\nabla}_j \left( f_{\tilde{S}} \right)^j_i = \alpha \gratio \left( L_i + \QQ u_{i} \right).
\end{align}

\subsection{Conservation of Energy}

To derive the energy equation, we begin by using \eqref{eq:vector_identity} again, applying it to the projection of \eqref{eq:enmom_conservation}
\begin{align*}
n_{\mu} \nabla_{\nu} T^{\mu \nu} - \nabla_{\mu} \left( \rho u^{\mu} \right) = n_{\mu} \QQ u^{\mu}
\end{align*}
where $n_{\mu} = \left(-\alpha, 0, 0, 0 \right)$ is the vector normal to each hypersurface. We can rewrite this as
\begin{align*}
\nabla_{\nu} \left( n_{\mu} T^{\mu \nu} - \rho u^{\nu} \right) = T^{\mu \nu} \nabla_{\nu} n_{\mu} - \alpha \QQ u^{0}.
\end{align*}
For convenience, we define the variable $\eta^{\nu} \equiv n_{\mu} T^{\mu \nu} - \rho u^{\nu}$, giving
\begin{align*}
\nabla_{\nu} \eta^{\nu} = T^{\mu \nu} \nabla_{\nu} n_{\mu} - \alpha \QQ u^{0}.
\end{align*}
Continuing in the same fashion as \ref{sec:lepton_baryon}, we arrive at
\begin{equation*}
    \partial_t \left(\alpha \gratio \eta^0 \right) + \hat{\nabla}_i \left( \alpha \gratio \eta^i \right) = \alpha \gratio \left( T^{\mu \nu} \nabla_{\nu} n_{\mu} - \alpha \QQ u^{0} \right).
\end{equation*}
After further calculations, we obtain
\begin{align*}
\partial_t \tilde{\tau} + \hat{\nabla}_i \left(f_{\tilde\tau}\right)^i = \alpha \gratio \left( \alpha \QQ u^{0} - T^{\mu \nu} \nabla_{\nu} n_{\mu} \right),
\end{align*}
with
\begin{align*}
    \tilde{\tau} &= \alpha^2 \gratio T^{00} - \tilde{D}, \\
    \left( f_{\tilde\tau} \right)^i &\equiv \gratio \left(\alpha^2 T^{0 i}\right) - \tilde{D} v^i,
\end{align*}
where $\tilde{\tau}$ is the conserved energy. We then expand the source terms, directing the reader again to~\cite{Montero_2014} for a detailed derivation. After additional work, we arrive at the the definition for the source term:
\begin{align}
    s_{\tilde\tau} &\equiv T^{00} \left( \beta_i \beta_j K^{ij} - \beta^i \partial_i \alpha \right) \nonumber \\ 
    &+ T^{0i} \left( 2 \beta_j \gamma_{ik} K^{jk} -  \partial_i \alpha \right) + T^{ij} K_{ij}, \nonumber
\end{align}
where $K^{ij}$ is the extrinsic curvature. Collecting all terms, we finally write: 
\begin{align}
    \partial_t \tilde{\tau} + \hat{\nabla}_i \left(f_{\tilde\tau}\right)^i = \alpha \gratio \left( s_{\tilde\tau} + \alpha \QQ u^{0} \right).
\end{align}

\subsection{Final Equations}

To conclude this section we rewrite our evolution equations in a modified version of the standard flux-conservative form, and set $\bar{\gamma} = \hat{\gamma}$. Our equations then take the form
\begin{equation}\label{eq:basic_evol}
\partial_{t} \consv + \hat{\nabla}_{i}\fluxv^{i} = \sourcev,
\end{equation}
and can be explicitly written as
%
\begin{widetext}
\begin{align}
    \renewcommand\arraystretch{1.4}
    \partial_{t}  
    \begin{bmatrix}
      \tilde{D}\\
      \yet\\
      \tilde{\tau}\\
      \tilde{S}_{j}
    \end{bmatrix}
    +
    \partial_{i}
     \begin{bmatrix}
        \big( f_{\tilde{D}} \big)^i\\
        \big( f_{\yet} \big)^i\\
        \big( f_{\tilde\tau} \big)^i\\
        \big( f_{\tilde{S}} \big)^i_j
    \end{bmatrix}
    =
    \begin{bmatrix}
        - \updowndowntensor{\hat{\Gamma}}{i}{i}{j} \big( f_{\tilde{D}} \big)^j\\
        \alpha e^{6\phi} \RR - \updowndowntensor{\hat{\Gamma}}{i}{i}{j} \big( f_{\yet} \big)^j\\
        \alpha e^{6\phi} \big( s_{\tilde\tau} + W \QQ \big) - \updowndowntensor{\hat{\Gamma}}{i}{i}{j} \big( f_{\tilde\tau} \big)^j\\
        \alpha e^{6\phi} \big(L_j + \QQ u_j \big) - \updowndowntensor{\hat{\Gamma}}{i}{i}{k} \big( f_{\tilde{S}} \big)^k_j + \updowndowntensor{\hat{\Gamma}}{k}{i}{j} \big( f_{\tilde{S}} \big)^i_k    
    \end{bmatrix},\;
    \label{eq:basic_evol_expanded}
\end{align}
where the conservative variables $\consv$ in \eqref{eq:basic_evol} are related to the primitive variables via
\end{widetext}

\begin{align*}
    \consv = 
    \begin{bmatrix}
        \tilde{D}\\
        \yet\\
        \tilde{\tau}\\
        \tilde{S}_{j}
    \end{bmatrix}
    &\equiv
    e^{6\phi}
    \begin{bmatrix}
        W\rho\\
        W\rho\ye\\
        \alpha^{2}T^{00} - W\rho\\
        \alpha T^{0}_{\ j}\\
    \end{bmatrix}.\;
\end{align*}
The conserved variables, fluxes $\fluxv$ and source terms $\sourcev$ are defined in terms of the primitive quantities \mbox{$\primv \equiv \left[ \rho, P, v^{i}, \ye, T \right]$}, where $T$ is the temperature. Note that our equations differ from those presented in~\cite{Montero_2014} due to the inclusion of lepton number conservation and neutrino emission. To summarize, transitioning from the formulation presented in~\cite{Werneck_IGM} to our approach using a reference metric for pure hydrodynamics involves:
\begin{enumerate}
    \item Replacing partial derivatives with covariant derivatives with respect to a reference metric.
    \item Replacing factors of $\sqrt{\gamma}$ with $e^{6\phi}$\;.
\end{enumerate}

\section{Numerical Implementation}
\label{num_int}

Our code employs a second-order accurate finite-volume scheme that approximates cell-averaged quantities using their cell-centered values. The \bhah infrastructure~\cite{Ruchlin_2018} provides a uniform grid discretization for coordinates $x^i = \left (x^1, x^2, x^3 \right)$ of the form
\begin{equation*}
    x^i \left( j \right) = \Delta x^i\left( j - N_G + \frac{1}{2} \right),
\end{equation*}
where $j$ is the grid index, $\Delta x^i$ is the grid spacing, and $N_G$ is the number of ghost cells, used for finite differences, interpolations, and reconstruction. In our implementation both the BSSN and GRHD quantities are defined on this same grid.

To second-order accuracy, cell-averages may be approximated by the \textit{volume centroid} in any given grid cell. While \textit{geometric cell-centers} and \textit{volume centroids} coincide in Cartesian coordinates, they differ in general curvilinear coordinates, thus cell-center values are generally not equivalent to their cell averages without corrections~\cite{Monchmeyer_1989, Mignone_2014}. However, \cite{Mewes_2020} demonstrated that this care need not be taken when working within a reference-metric formulation of the GRHD equations, since their integral form is effectively ``Cartesian''. In short, because we reformulate the equations with respect to the background metric $\hat{\gamma}_{ij}$ and express them in an \textit{orthonormal basis}, as shown in the following subsection, information about the underlying coordinate system is transferred to the source terms. In fact, if we take volume integrals over \eqref{eq:basic_evol_expanded}, in the \textit{orthonormal basis} of our chosen coordinate system, $\hat{\gamma}_{ij}$ simplifies to the flat metric and the integrands do not acquire additional geometric scale factors. This is unlike~\cite{Monchmeyer_1989, Mignone_2014}, where all scale factors appear in the integrands. We refer the reader to~\cite{Mewes_2020} for further discussion. In the following subsections we outline key details of our numerical scheme. 

\subsection{Rescaling Tensorial Quantities}

Numerical integration of~\eqref{eq:basic_evol_expanded} in singular coordinate systems often leads to instability as tensor components can diverge or vanish at coordinate singularities. To mitigate this, we follow the approach of~\cite{Montero_2014, Mewes_2020}, employing cell-centered grids and integrating the equations in an \textit{orthonormal basis}, which removes singular terms from tensors subject to evolution or finite differencing.

To implement this strategy, we define the rescaling quantities:
\begin{equation}
    \mathcal{R}_i = \sqrt{\hat{\gamma}_{ii}}, \quad 
    \mathcal{R}^i = \frac{1}{\mathcal{R}_i}, \quad 
    \mathcal{R}^{ij} = \mathcal{R}^i \odot \mathcal{R}^j,
    \label{eq:rescaling_quantities}
\end{equation}
where $\odot$ denotes the Hadamard product, indicating element-wise multiplication (without summation over repeated indices).

As an example, when computing the finite difference of the flux term for the densitized momentum, we first rescale the flux tensor $\updowntensor{F}{j}{i} \equiv \alpha \gratio \updowntensor{T}{j}{i}$ as follows:
\begin{equation}
    \updowntensor{f}{j}{i} = \updowntensor{F}{j}{i} \odot \mathcal{R}_j \odot \mathcal{R}^i.
    \label{eq:rescale_flux}
\end{equation}
This rescaling ensures that if the fluxes are smooth in Cartesian coordinates, the rescaled flux tensor $\updowntensor{f}{j}{i}$ remains smooth across coordinate singularities. The flux divergence $\partial_j \updowntensor{F}{j}{i}$ is then expressed in terms of $\updowntensor{f}{j}{i}$ as:
\begin{equation}
    \partial_j \updowntensor{F}{j}{i} = \partial_j \left( \updowntensor{f}{j}{i} \odot \mathcal{R}_i \odot \mathcal{R}^j \right).
    \label{eq:rescaled_flux_divergence}
\end{equation}
In practice, we evaluate \eqref{eq:rescaled_flux_divergence} using the product rule, taking finite differences of $\updowntensor{f}{j}{i}$ and computing derivatives of $\mathcal{R}^j$ analytically. Since the conserved momentum is a contravariant vector containing singular terms (first appearing from the transformation of initial data from Cartesian to curvilinear coordinates), we scale out these terms to avoid numerical issues. Thus, we evolve $\tilde{s}_i = \tilde{S}_i \odot \mathcal{R}^i$ in the Euler equation, and all GRHD equations are expressed in terms of the \textit{rescaled} three-velocity. A similar rescaling procedure is applied to integrated, but not finite-differenced, quantities in the reference metric formulation of the BSSN equations, as discussed in~\cite{Baumgarte_2013, Ruchlin_2018}. This ensures consistency across both finite-differenced and integrated terms, mitigating issues arising from coordinate singularities.

\subsection{High-Resolution Shock Capturing Scheme}

\groovy is built on the \nrpy/\bhah framework~\cite{Ruchlin_2018,Etienne_nrpytutorial}, which provides the core infrastructure for evolving the coupled matter and spacetime fields. The evolution equations are expressed symbolically, enabling \nrpy to automatically generate optimized C code for all tensorial and finite-differenced equations. Written in covariant form with respect to a reference metric, these equations can be generated in any coordinate system at \nrpy code-generation time with a single parameter choice. 

While future applications will leverage \bhah's multi-patch infrastructure, the current results solve the GRHD and Einstein's equations on a single patch. \bhah provides various time-stepping algorithms and both extrapolation and radiation boundary conditions. For all tests in Section \ref{code_tests}, we use RK4 time-stepping and apply Sommerfeld boundary conditions to the spacetime variables, while primitive variables are copied from the grid interior to the outer boundaries. 

To handle complex boundary conditions in curvilinear coordinates---particularly near angular boundaries and the origin---we use \bhah's native curvilinear boundary condition driver. This driver automatically addresses parity changes caused by basis vector sign flips across coordinate boundaries. Since \bhah employs cell-centered grids, ghost cells at inner boundaries (e.g., $r<0$ or $\theta>\pi$) directly overlap interior grid cells, requiring only mapping and accounting for basis vector direction changes. For further details, see~\cite{Baumgarte_2013, Ruchlin_2018, Etienne_nrpytutorial, Mewes_2020}.

To close the evolution equations in \eqref{eq:basic_evol_expanded}, we incorporate an equation of state (\eos) via direct linkage with \grhayl, which performs point-wise \eos computations. \grhayl supports various \eos models, including tabulated and hybrid piecewise polytropic models, as well as a simple ideal gas model that does not distinguish between cold and thermal components.

After each sub-step of the time evolution scheme, we recover the primitive variables from the conservative variables, typically using a non-linear solver. \grhayl provides several widely used conservative-to-primitive solvers, including the two-dimensional Noble2D solver from~\cite{Noble_2006} and the one-dimensional solvers from~\cite{Font_2000, Palenzuela_2015, Newman_2014}. Backup options are employed to maintain physical realism in cases where inversions fail, particularly in the low-density atmosphere, where truncation errors in the conservative variable evolution can push values outside physical bounds. Since these routines operate in a Cartesian basis, we transform the inputs to Cartesian coordinates before invoking the solver and convert the outputs back to the original coordinate system afterward.

Time evolution of the conservative variables is governed by \eqref{eq:basic_evol_expanded}, with flux terms computed using the Harten-Lax-van Leer (HLL) approximate Riemann solver~\cite{Harten_1983}. As in most fluid-based codes with high-resolution shock-capturing schemes, we interpolate metric quantities and primitive variables to cell interfaces before calculating the HLL fluxes. For primitive variable reconstruction, we use \grhayl's routines, employing either the piece-wise linear method~\cite{TVD} or the piece-wise parabolic method (PPM)~\cite{PPM1}. We apply reconstruction to the \textit{rescaled} three-velocity to prevent the PPM shock detection algorithm from misinterpreting coordinate singularities as physical shocks. While modifications to PPM for curvilinear coordinates have been explored in past work~\cite{Mignone_2014, Felker_2018}, their connection to our rescaling technique remains unclear. A close examination of the PPM algorithm reveals that the scheme considers gradients of the pressure, which may be the cause of the undesirable results noted by~\cite{Mewes_2020}, as PPM reconstruction without modification will yield incorrect values for the pressure gradient without additional geometric terms. We have observed similar issues in our implementation, as discussed in Section \ref{uniform_rotation}.

Finally, we incorporate the \texttt{NRPyLeakage}~\cite{Werneck_IGM} neutrino physics module from \grhayl, enabling our code to model neutrino emission via a leakage scheme. The specific reactions included in this scheme are detailed in~\cite{Werneck_IGM}. At each time step, the module processes the primitive variables and outputs the numerical values of $\RR$ and $\QQ$, as defined in \eqref{eq:lepton_number_conservation} and \eqref{eq:enmom_conservation}.

\subsection{Algorithmic Differences between \groovy and \igm}

\igm is a GRMHD code within the \etk framework, designed for modeling magnetized fluids in dynamical spacetimes with neutrino leakage. Building on \igm's core algorithms, \groovy uses \grhayl to implement GRHD capabilities within the \bhah infrastructure. While preserving most of \igm's original implementation choices, \groovy introduces two key enhancements aimed at improving accuracy and physical fidelity.

First, unlike the original \igm implementation, which interpolates BSSN quantities to \textit{cell-interfaces} at third-order accuracy before applying second-order finite differencing to approximate source terms at \textit{cell-centers}, \groovy directly operates on cell-centered quantities. This approach eliminates interpolation errors in spacetime quantities used to compute the GRHD source terms, and enables the use of any finite-differencing order supported by \nrpy, currently up to 12th order.

The advantages of this approach are detailed in Appendix~\ref{appendix:finite_difference_study}, where we demonstrate that increasing the finite-differencing order for metric source terms from second to fourth order significantly reduces central density drift in equilibrium neutron star simulations. Beyond fourth order (e.g., sixth order), the improvements are negligible; therefore, \groovy adopts fourth-order finite differencing for metric derivatives in GRHD source terms. Given these results, we plan to incorporate higher-order finite-difference schemes for metric derivative source terms into future versions of \igm.

Second, in evolving initially cold hydrodynamic flows modeled by polytropic or piecewise polytropic \eoss, fluid pressures can deviate from their cold values only through shock heating; there is no mechanism for cooling below the initial cold pressure $P_{\rm cold}$. However, numerical errors can drive $P < P_{\rm cold}$, especially near neutron star surfaces. In Appendix~\ref{appendix:pressure_floor_study}, we show that applying a pressure floor set to the physical minimum, $P_{\rm cold}$, leads to larger central density drifts. Conversely, allowing unphysical pressures by decreasing the floor minimizes these drifts. The original \igm mitigated this issue by setting the floor to $0.9 P_{\rm cold}$, balancing diffusion rates against the allowance of unphysical pressures. In contrast, \groovy and \grhayl adopt a stricter approach, setting the pressure floor to the minimum value allowed by the physics, $P_{\rm cold}$.

\section{Code Tests}
\label{code_tests}

In the following subsections, we demonstrate our code's ability to solve standard numerical tests from the literature in both fixed (\ref{code_tests:flat_static}, \ref{code_tests:curved_static}) and dynamical (\ref{code_tests:dynamical}) spacetimes.

\subsection{Flat, Static Spacetime Tests}
\label{code_tests:flat_static}

\subsubsection{Balsara 0 Shock Test}

Standard MHD tests include the Balsara tests~\cite{Balsara_2001}, which evolve shocks and contact discontinuities in a magnetized plasma in 1D, typically using a $\Gamma$-law \eos. To showcase the high-resolution shock-capturing capabilities of our implementation in curvilinear coordinates, we evolve the Balsara 0 test (the Balsara 1 test without magnetic fields) in spherical coordinates. The initial conditions for the stationary neutral plasma are:
\begin{align*}
\rho\left( z \right) &= \left \{ \begin{array}{lll}
1.0 & \mbox{;} & z \leq z_0 \\
0.125 & \mbox{;} & z > z_0 \end{array} \right\}, \\
P\left( z \right) &= \left \{ \begin{array}{lll}
1.0 & \mbox{;} & z \leq z_0 \\
0.1 & \mbox{;} & z > z_0 \end{array} \right\},
\end{align*}
where $z_0$ is the initial shock location. Unlike previous studies~\cite{Spritz, Balsara_2001}, which place the shock at the origin, we set $z_0 = 1.0$, propagating the shock along the $z$-axis, transforming the 1D Cartesian test into a significantly more challenging 2D spherical-coordinate test.

We set the outer boundary at $r_{\text{max}} = 2.0$ and evolve the data to $t = 1.0$ using total variation diminishing (TVD) reconstruction with a minmod limiter~\cite{TVD} and the Noble2D hybrid \eos conservative-to-primitive solver from \grhayl. The resolution is $\left( N_r, N_\theta, N_\phi \right) = \left( 520, 260, 2 \right)$, and we adopt a $\Gamma$-law \eos of the form $P = (\Gamma - 1) \rho \epsilon$, where $\epsilon$ is the specific internal energy and $\Gamma = 2$. This setup uses the simple \eos option from \grhayl's \eos module.

\begin{figure}[ht]
  \centering
  \includegraphics[width=\linewidth]{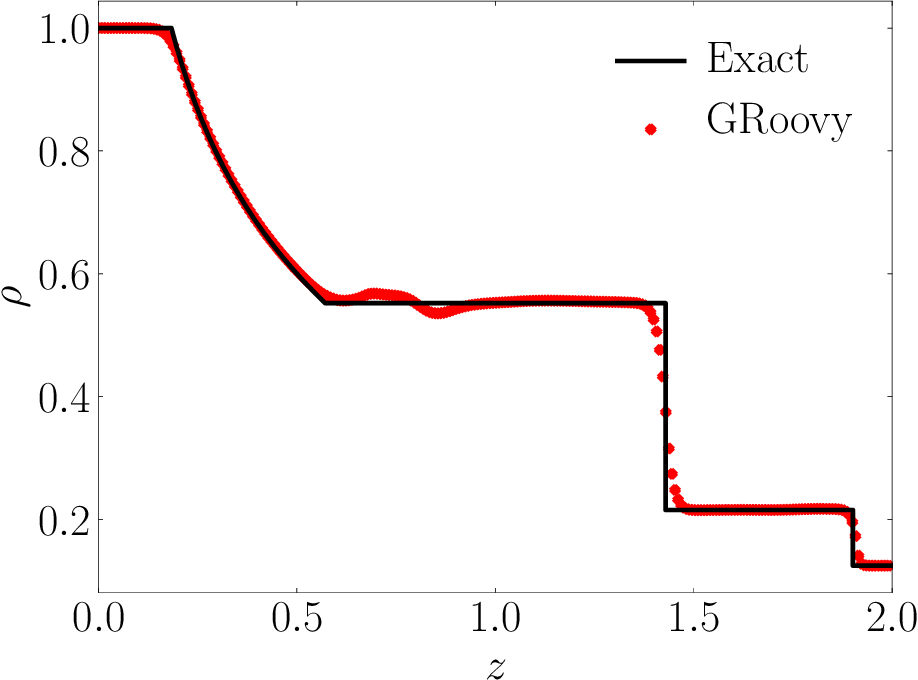}
  \caption{Density profile at time $t=1.0$ from the evolution of the Balsara 0 initial data.}
  \label{fig:balsara0}
\end{figure}

In \figref{fig:balsara0}, we compare our results with the exact solution obtained using the Riemann solver from~\cite{Giacomazzo_2006}. While our results show good agreement, small oscillations are present, likely caused by the uneven resolution along the shock front as the shock and contact discontinuity propagate in opposite directions along the $z$-axis in spherical coordinates. This demonstrates that our code can successfully handle standard numerical tests in more challenging curvilinear coordinate settings.

\subsubsection{Spherical Explosion}

\begin{figure*}[!ht]
  \centering
  \includegraphics[width=\textwidth]{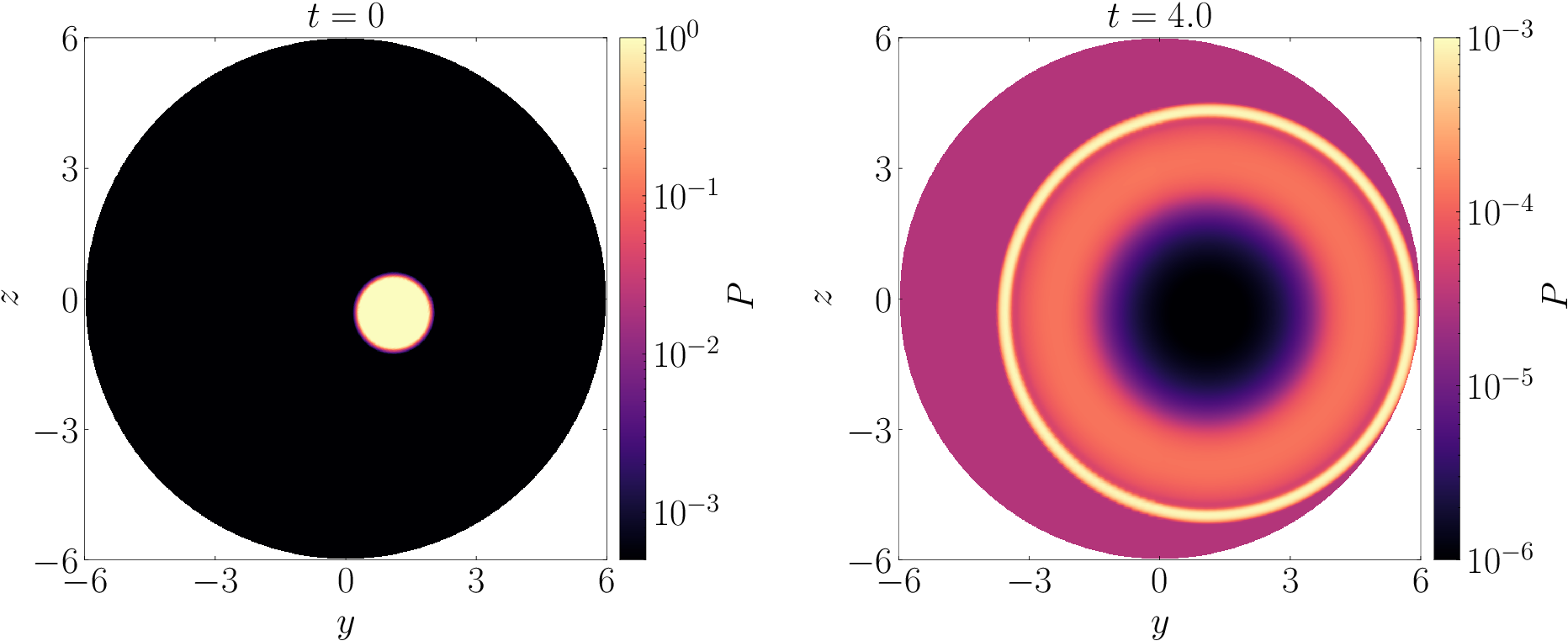}
  \caption{Pressure profiles from the spherical explosion test, at $t=0$ and $t=4.0$.}
  \label{fig:sph_exp}
\end{figure*}

To demonstrate our code's ability to handle hydrodynamic flows across coordinate singularities, we perform a standard hydrodynamic test: the spherical explosion~\cite{Komissarov_1999}. This test models an over-dense ball of gas expanding into an ambient medium, a scenario most efficiently evolved in spherical coordinates with the explosion centered at the origin. It becomes a ``torture'' test when the explosion is offset from the $z$-axis (as in~\cite{Mewes_2020}), requiring the code to handle hydrodynamic propagation across coordinate singularities while preserving spherical symmetry. This makes it a valuable benchmark for GRMHD and GRHD codes in spherical coordinates.

Following~\cite{Mewes_2020}, we evolve the initial data in spherical coordinates, offsetting the explosion profile by $y = 1.1$, and additionally by $z = 0.3 \times y$. The initial density profile, following~\cite{Spritz}, is given by:
\begin{align}
\rho\left( r \right) &= \left \{ \begin{array}{lll} \rho_{\rm in} & \mbox{;} & r \leq r_{\rm in} \\
\exp \frac{\left( r_{\rm out} - r \right) \ln\rho_{\rm in} + \left( r - r_{\rm in} \right) \ln\rho_{\rm out}}{r_{\rm out} - r_{\rm in}} & \mbox{;} & r_{\rm in} < r < r_{\rm out} \\ 
\rho_{\rm out} & \mbox{;} & r \geq r_{\rm out} \end{array} \right.
\end{align}
A similar pressure profile is used, where $r_{\rm in}$ and $r_{\rm out}$ denote the inner and outer radii, respectively, with $r_{\rm in} = 0.8$ and $r_{\rm out} = 1.0$. The inner region has $\rho_{\rm in} = 10^{-2}$ and $p_{\rm in} = 1.0$, while the outer region has $\rho_{\rm out} = 10^{-4}$ and $p_{\rm out} = 3 \times 10^{-5}$. We adopt a $\Gamma$-law \eos with $\Gamma = 4/3$, use monotonized central (MC) reconstruction~\cite{TVD}, and set the grid resolution to $\left( N_r, N_\theta, N_\phi \right) = \left( 160, 80, 160 \right)$.

We observe that the shock front successfully propagates through the $z$-axis coordinate singularity without significant artifacts, as shown in \figref{fig:sph_exp}. While the shock front slows down artificially as the radial velocity component crosses zero at the singularity, this slowdown has no lasting effects at later times.

\subsubsection{Neutrino Tests}

\begin{figure*}[!ht]
    \captionsetup[subfigure]{labelformat=empty}
        \begin{subfigure}{.475\textwidth}
            \centering
            \includegraphics[width=\textwidth]{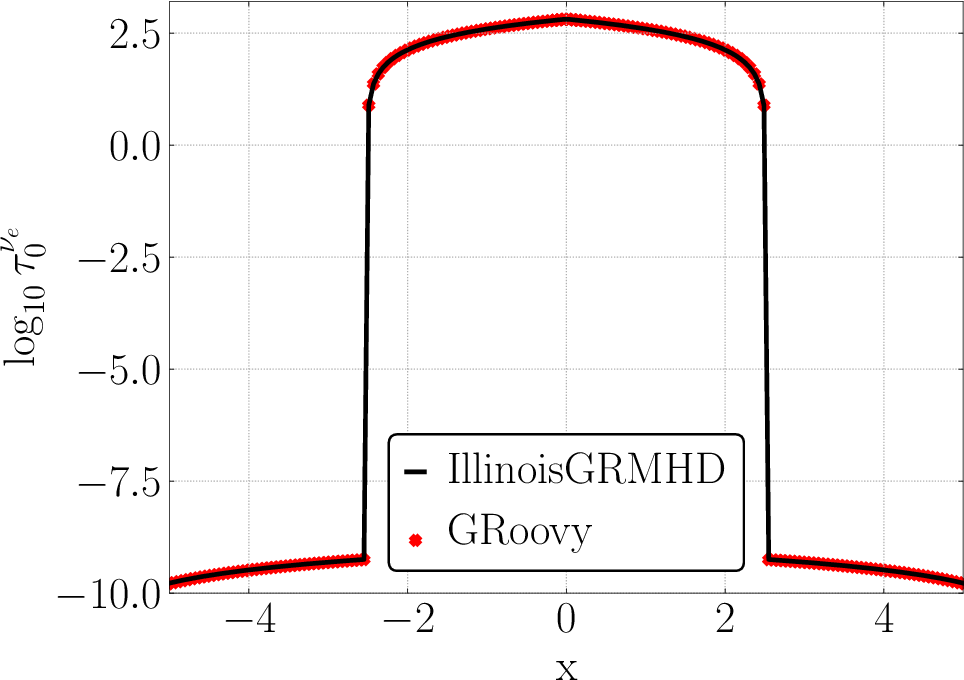}
            \label{fig:neutrinos_cons_dens}
    \end{subfigure}%
    \hspace{1em}
        \begin{subfigure}{.475\textwidth}
            \centering
            \includegraphics[width=\textwidth]{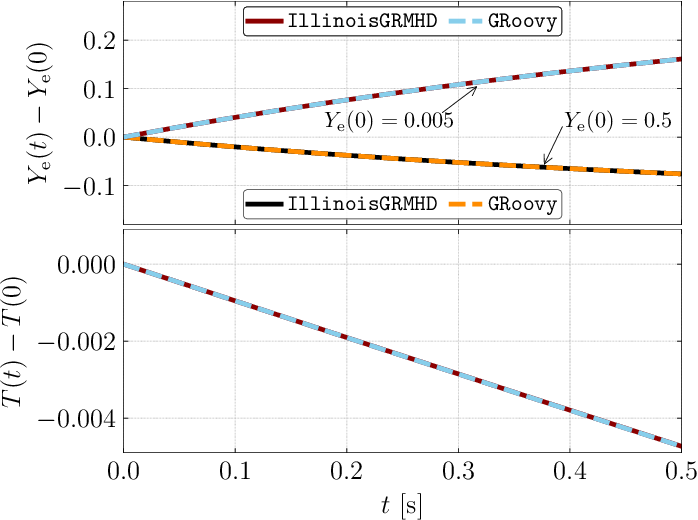}
            \label{fig:neutrinos_isogas}
    \end{subfigure}%
    \caption{\textbf{Left}: Radial profile of optical depth from the initialization procedure for an optically thick gas. \textbf{Right}: Time evolution of the electron fraction $\ye$ and temperature $T$ for an optically thin gas.}
    \label{fig:neutrinos_overall}
\end{figure*}

We conclude our flat spacetime tests with a set of neutrino tests, demonstrating the successful integration of \texttt{NRPyLeakage}~\cite{Werneck_IGM} into our code via \grhayl's neutrinos module.

The first test evaluates the code's ability to initialize optical depth in a constant-density sphere, representing an optically thick gas. The sphere has a radius of $r_{\text{Sph}} = 2.5$, density $\rho^{\text{Sph}} = 1.58 \times 10^{-4}$, electron fraction $\ye^{\text{Sph}} = 0.1$, and temperature $T^{\text{Sph}} = 8.0$, embedded in an ambient medium with $\rho^{\text{Ext}} = 9.71 \times 10^{-11}$, $\ye^{\text{Ext}} = 0.5$, and $T^{\text{Ext}} = 0.01$. For this test, we use the SLy4 \eos~\cite{Schneider_2017} and a spherical coordinate grid with resolution $\left( N_r, N_\theta, N_\phi \right) = \left( 60, 2, 2 \right)$. After applying this algorithm, we find excellent agreement between the optical depth obtained from our initialization procedure and the result from \grhaylmhd~\cite{Werneck_IGM}, as shown in the left panel of \figref{fig:neutrinos_overall}.

The second test of our implementation of \texttt{NRPyLeakage} models an optically thin, neutrino-laden isotropic gas in spherical coordinates. We evolve two versions of the initial data, with $\ye\left(0 \right) = 0.5$ and $\ye\left(0 \right) = 0.005$. For both tests, we set $\rho = 10^{-12}$ and $T = 1$. In the right panel of \figref{fig:neutrinos_overall}, we compare our results against \grhaylmhd, confirming agreement between the solutions.

\subsection{Curved, Static Spacetime Tests}
\label{code_tests:curved_static}

\subsubsection{Non-Rotating NS with a Hybrid \eos}
\label{code_tests:static_spacetime_hybrid_eos_tov}

An essential test for GRHD codes is the stable evolution of an isolated neutron star, where hydrodynamic fields are coupled to a curved spacetime. This test, often performed in Cartesian coordinates with adaptive mesh refinement (AMR), involves first setting up Tolman–Oppenheimer–Volkoff (TOV) initial data. The TOV equations provide a solution to Einstein's equations for a spherically symmetric star in hydrostatic equilibrium, but do not assume an \eos. To this end, we choose a cold polytropic \eos of the form
\begin{equation*}
    P = K\rho^\Gamma,
\end{equation*}
where $K$ is the polytropic constant and $\Gamma$ is the adiabatic index. For all polytropic \eos tests in this paper, we set a central density $\rho_c = 1.28 \times 10^{-3}$, $K=100$, and $\Gamma=2$. These parameters correspond to a star with a gravitational mass of $M = 1.4$ and a radius of $R = 8.1$ in code units, representing cold, degenerate nuclear matter. We generate these initial data using the TOV solver within \nrpy~\cite{Etienne_nrpytutorial}.

\begin{figure*}[ht!]
  \centering
  \includegraphics[width=0.49\linewidth]{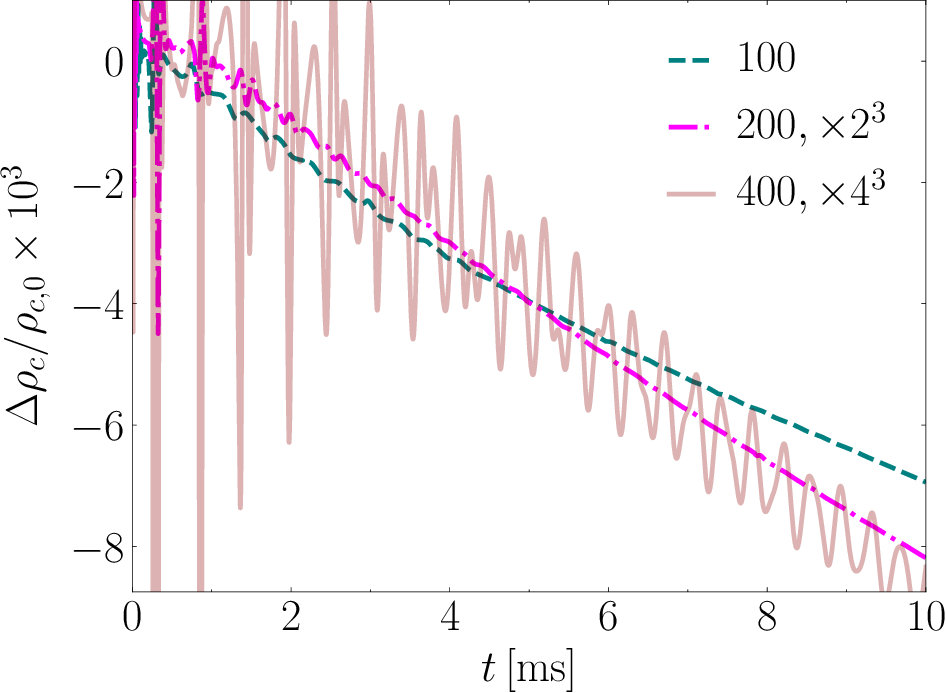}
  \includegraphics[width=0.49\linewidth]{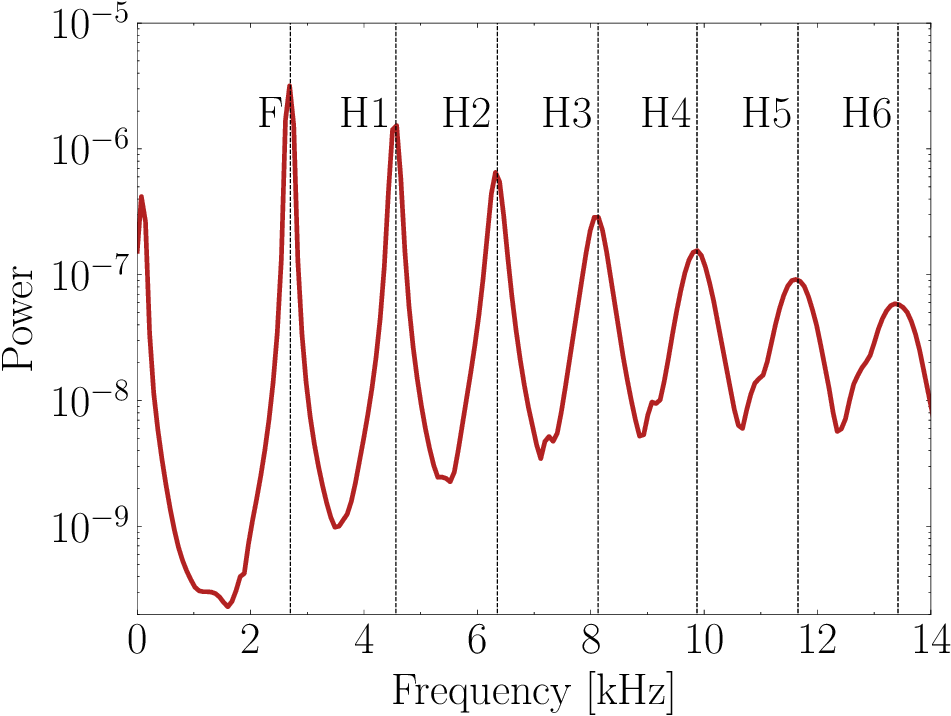}
  \caption{Static-spacetime TOV evolution in spherical coordinates: central density drift. \textbf{Left}: Convergence study showing approximately third-order convergence, with radial resolutions $N_r = \left(100, 200, 400\right)$ and angular resolutions fixed at $N_\theta = N_\phi = 2$. The high-resolution run captures more power in the high-frequency overtones, as illustrated in the right panel. \textbf{Right}: Power spectrum of the central density for $N_r=400$. Dashed vertical lines mark the fundamental mode (F) and overtones (H1--H6) from~\cite{Font_2002}. A Hann window is applied to the time series before performing the Fourier transform.}
  \label{fig:Cowling_TOV}
\end{figure*}

We test our code's ability to evolve these hydrodynamic fields atop a static spacetime---the Cowling approximation---in spherical coordinates. For these tests, we evolve the star with a hybrid \eos (with zero thermal part initially), set the outer boundary at $r_\text{max} = 20$, and maintain a constant-density atmosphere with $\rho_{\text{atm}} = \rho_c \times 10^{-13}$.

For this test, we use the Noble2D conservative-to-primitive solver with the 1D solver from~\cite{Palenzuela_2015} as a backup, and employ PPM reconstruction. Although the Noble2D solver occasionally fails to converge in low-density regions during the initial iterations, the 1D solver reliably resolves these cases, allowing the simulation to proceed smoothly. Finally, place the star at the origin, effectively making this a 1D test.

Since we employ a finite volume scheme to solve the GRHD equations, truncation errors induce an initial perturbation in the star, leading to oscillations. These perturbations arise both at the surface, where the density drops by orders of magnitude, and at the center, where PPM reduces to lower order at extrema~\cite{Duez_IGM}. These errors also cause the star to gradually diffuse over time, resulting in a downward drift of the central density. The left panel of \figref{fig:Cowling_TOV} demonstrates that doubling the resolution from $N_r=100$ to 200, and then from 100 to 400, reduces the diffusion by approximately a factor of 8 and 64, respectively. This result implies nearly third-order convergence, consistent with the expected convergence order of \grhaylmhd, which has been shown to achieve accuracy between second and third order.

Meanwhile, the magnitude of oscillations induced at the start of the evolution increases significantly as the radial resolution is increased from 200 to 400--a much higher radial sampling than is typical in numerical relativity simulations, with over 320 points across the diameter of the star. This behavior, as shown in the right panel of \figref{fig:Cowling_TOV}, results from higher harmonics from the initial perturbation being better resolved and not diffused away. When run at high resolution, our code successfully maintains power up to the sixth overtone for this TOV model, consistent with the frequencies reported in~\cite{Font_2002}.

\subsubsection{Non-Rotating NS with a Tabulated \eos and Neutrino Leakage}
\label{code_tests:cowling_spacetime_tab_eos_tov}

Hybrid \eos prescriptions have been widely adopted in the literature with considerable success, but they lack the capability to accurately model finite temperature effects compared to more realistic \eoss. In this section we model a TOV solution using the SHT~\cite{Shen_2011} \eos. We reproduce the sTOV-SHT model from~\cite{Galeazzi:2013mia} using the solver in \nrpy, constructing the initial data from a beta-equilibrium, constant-entropy slice of the \eos table, at specific entropy $s = 1\,k_B$. The resulting solution has a central density of $9.3 \times 10^{14} \mathrm{g\,cm}^{-3}$, a gravitational mass of $2.741\,M_{\odot}$, and a central temperature of $\sim30\,\mathrm{MeV}$. We verified the correctness of our optical depth calculation for this case as well, finding qualitative agreement with Fig.~1 of~\cite{Neilsen:2014hha}.

\begin{figure*}[ht!]
 \centering
 \includegraphics[width=0.49\linewidth]{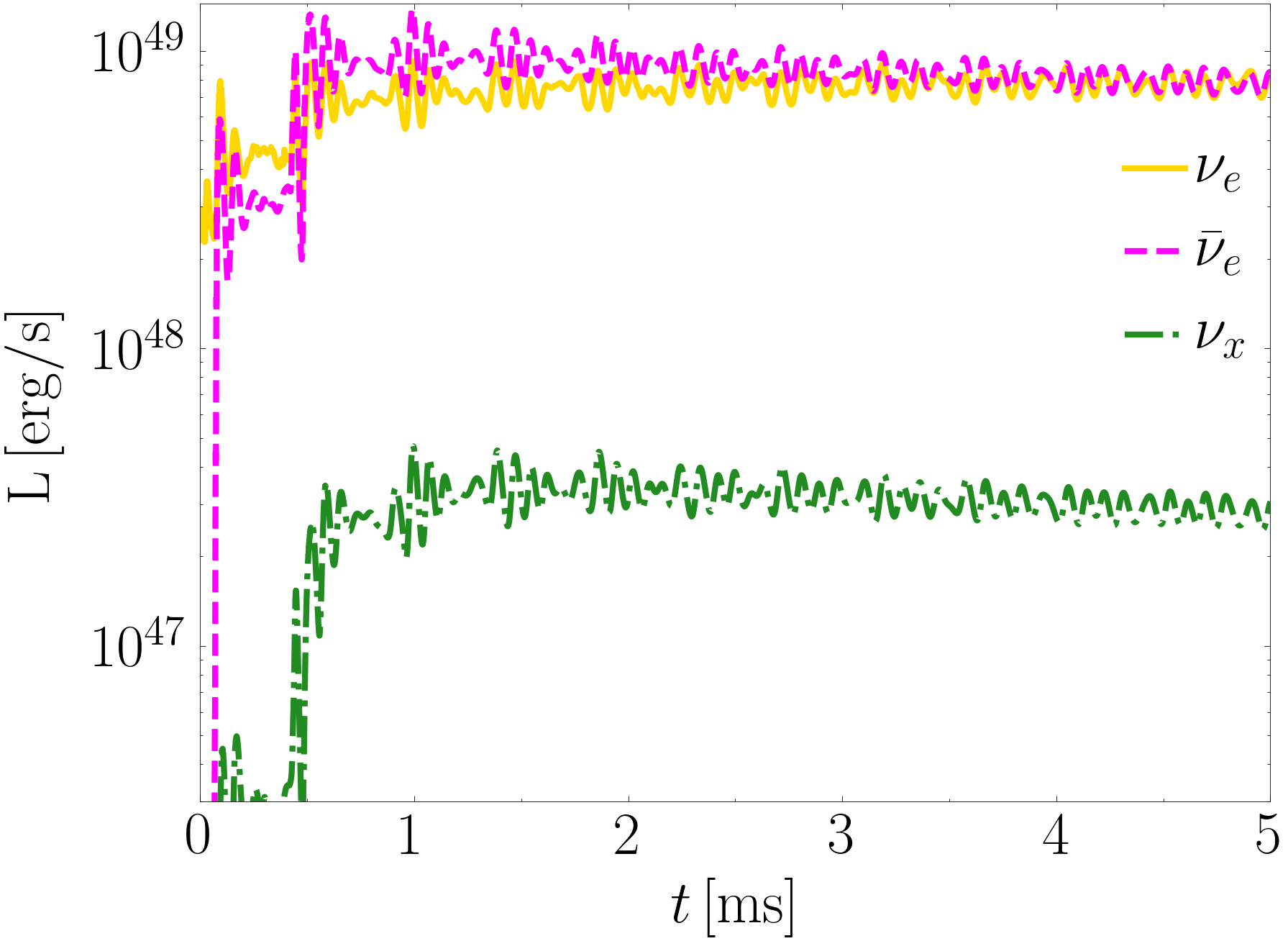}
 \includegraphics[width=0.49\linewidth]{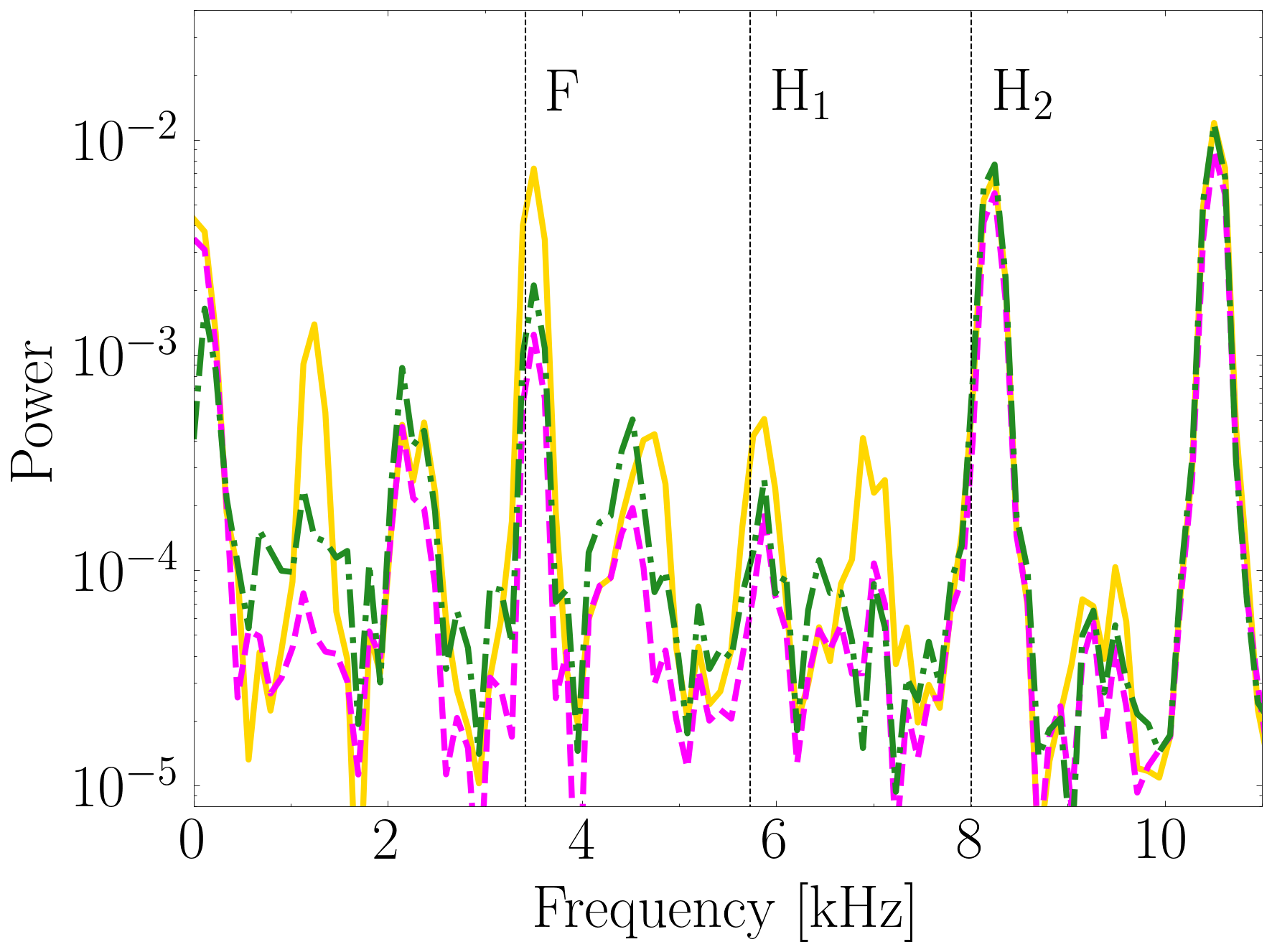}
 \caption{Luminosities computed using the neutrino leakage module of \grhayl, from evolving a hot TOV model using the SHT~\cite{Shen_2011} \eos in a static spacetime. \textbf{Left}: Time evolution of neutrino luminosities, using spherical coordinates with a radial resolution of $25\,\mathrm{m}$. \textbf{Right}: Normalized power spectrum for all three neutrino species, showing the fundamental mode and overtones. Reference frequencies taken from~\cite{Galeazzi:2013mia}.}
 \label{fig:Cowling_Leakage_TOV}
\end{figure*}

For the test presented here, we use spherical coordinates, PPM reconstruction, enable neutrino leakage, and use the one-dimensional conservative-to-primitive routine of~\cite{Newman_2014} as implemented in \grhayl. To compare our results with those found in~\cite{Neilsen:2014hha} using the \code{GR1D}~\cite{OConnor:2009iuz} code, which solves the GRHD equations in spherical symmetry, we use a radial resolution of $25\,\mathrm{m}$, and place the outer boundary at $r_{\mathrm{max}} = 62\,\mathrm{km}$. Finally, we use a constant density atmosphere set to the minimum value of density in the \eos table. We show our results in \figref{fig:Cowling_Leakage_TOV}, plotting the time evolution of the luminosities of electron neutrinos $\nu_e$, electron antineutrinos $\bar{\nu}_e$, and heavy lepton neutrinos $\nu_x = \left\{ \nu_{\mu},\nu_\tau \right\}$, in the left panel. We compute neutrino luminosities using
\begin{equation*}
   \mathrm{L}_{\nu_i} = \int \alpha^2 W Q^\mathrm{eff}_{\nu_i} \sqrtgamma d^3x,
\end{equation*}
where $Q^\mathrm{eff}_{\nu_i}$ is the effective emission rate for each $\nu_i = \left\{ \nu_e, \bar{\nu}_e, \nu_x \right \}$. In the right panel we show the dominant frequencies in the luminosity oscillations, finding good agreement with the frequencies quoted in~\cite{Galeazzi:2013mia}. We note that while our luminosities are three orders of magnitude lower than those found in~\cite{Galeazzi:2013mia} and about two orders lower than what~\cite{Neilsen:2014hha} found, both studies determined that these luminosities are strongly dependent on the atmosphere prescription. Thus, we attribute this result to the low numerical viscosity in the code when using spherical coordinates. This point will be expanded upon in the following section. In our testing we found that setting the atmospheric density to approximately $\rho_{\text{atm}} = \rho_c \times 10^{-6}$ leads to an increase in the luminosities by about two orders of magnitude, and they increase further when using cylindrical coordinates. Our results also seem reasonable given that~\cite{Galeazzi:2013mia} noted that the luminosities extracted from evolving an initially cold model, reaching a maximum of $\sim 10^{48} \mathrm{erg\,s}^{-1}$, should be treated as the minimum luminosity their leakage scheme can model.

\subsection{Curved, Dynamical Spacetime Tests}
\label{code_tests:dynamical}

\subsubsection{Non-Rotating NS with a Hybrid \eos}
\label{code_tests:dynamical_spacetime_hybrideos_tov}

\begin{figure*}[ht!]
  \centering
  \includegraphics[width=0.49\linewidth]{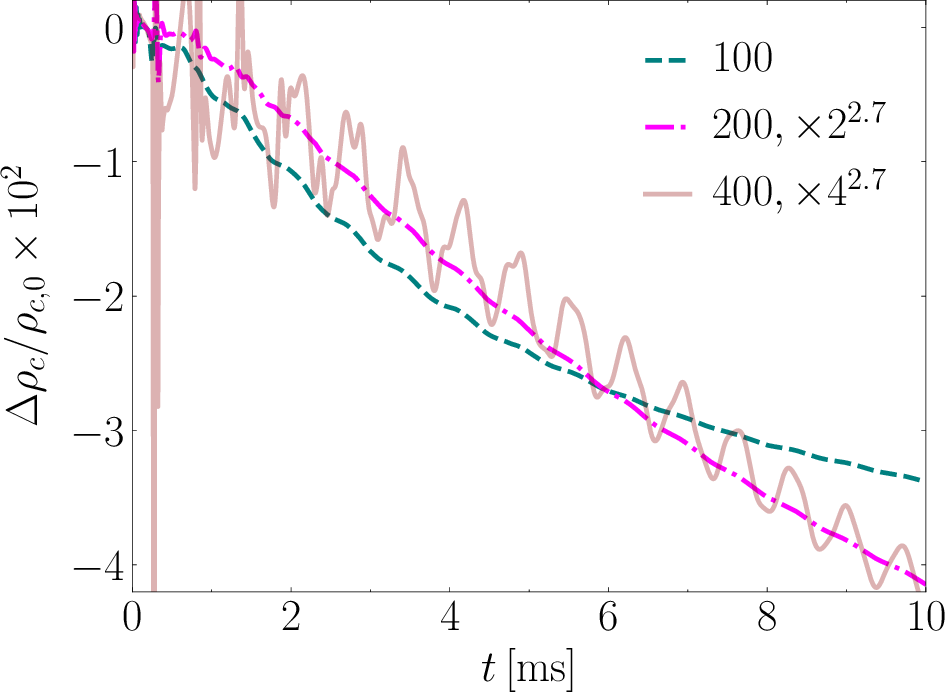}
  \includegraphics[width=0.49\linewidth]{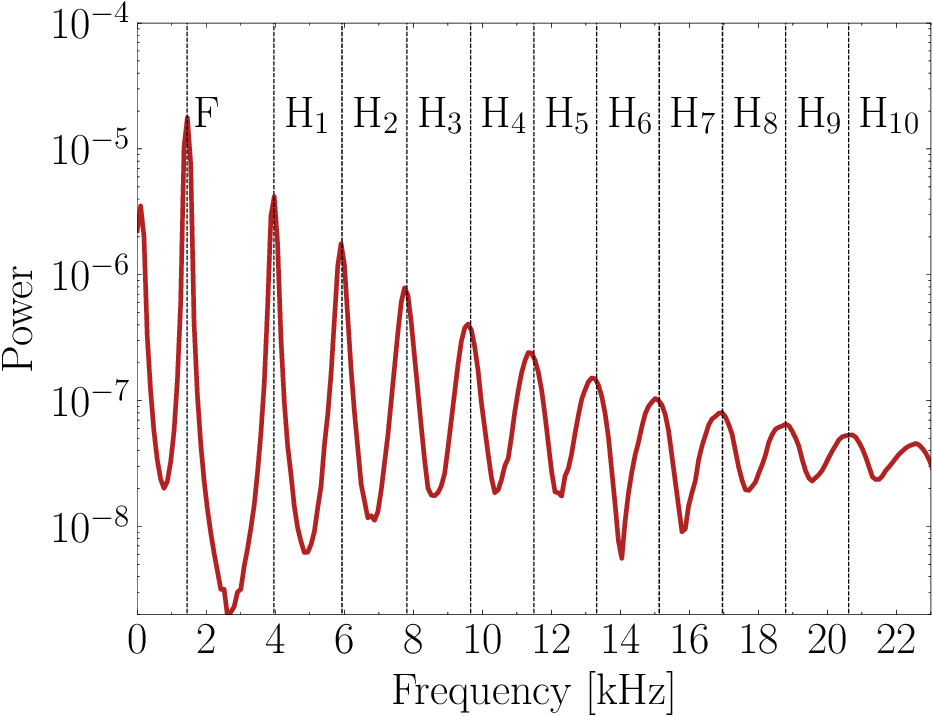}
\caption{Dynamical-spacetime TOV evolution in spherical coordinates: central density drift. \textbf{Left}: Similar to the left panel of \figref{fig:Cowling_TOV}, but with the vertical axis (density drift) rescaled by $10^2$ instead of $10^3$, and results at different radial resolutions rescaled assuming $2.7$-order convergence. \textbf{Right}: Similar to the right panel of \figref{fig:Cowling_TOV}, but including overtones up to the tenth harmonic provided by~\cite{Kokkotas_email}.}
  \label{fig:Dynamical_TOV}
\end{figure*}

Having demonstrated that \groovy performs as expected in both flat and curved-but-static spacetime tests, we now turn to dynamical spacetime tests, which involve coupling to our spacetime solver. For spacetime evolution, we utilize the BSSN formulation with moving-puncture gauge conditions, incorporating the $1+\log$ lapse evolution~\cite{Campanelli_2006, Baker_2006} and a second-order, non-covariant advecting shift evolution~\cite{Meter_2006}, governed by:
\begin{align*}
\partial_t \alpha &= \beta^i \partial_i \alpha - 2 \alpha K, \\
\partial_t \beta^i &= \beta^j \partial_j \beta^i + B^i, \\
\partial_t B^i &= \beta^j \partial_j B^i + \frac{3}{4} \partial_{0} \bar{\Lambda}^i - \eta B^i,
\end{align*}
where $K_{ij}$ is the extrinsic curvature, $K$ is its trace, $B^i$ and $\bar{\Lambda}^i$ are auxiliary evolution variables, and $\eta$ is a damping parameter. We use fourth-order finite differencing to represent spatial derivatives in the BSSN equations, set $\eta=0$, and apply Kreiss-Oliger dissipation~\cite{Kreiss_1973} with strength $k=0.2$, utilizing a fifth-order finite difference stencil.

In the left panel of \figref{fig:Dynamical_TOV}, we present the results of our convergence study for this test, using the same resolutions and initial data as \secref{code_tests:static_spacetime_hybrid_eos_tov}. The results are consistent with those from the static spacetime tests, showing that the drift in central density converges within the expected range (between second- and third-order), at approximately 2.7th order. The Fourier transform of the $N_r=400$ central density (right panel of \figref{fig:Dynamical_TOV}) shows frequencies in excellent agreement with predictions from a code deriving these frequencies from linear perturbations~\cite{Kokkotas_email}, extending up to the tenth overtone.

\begin{figure*}[!ht]
  \centering
  \includegraphics[width=1.0\textwidth]{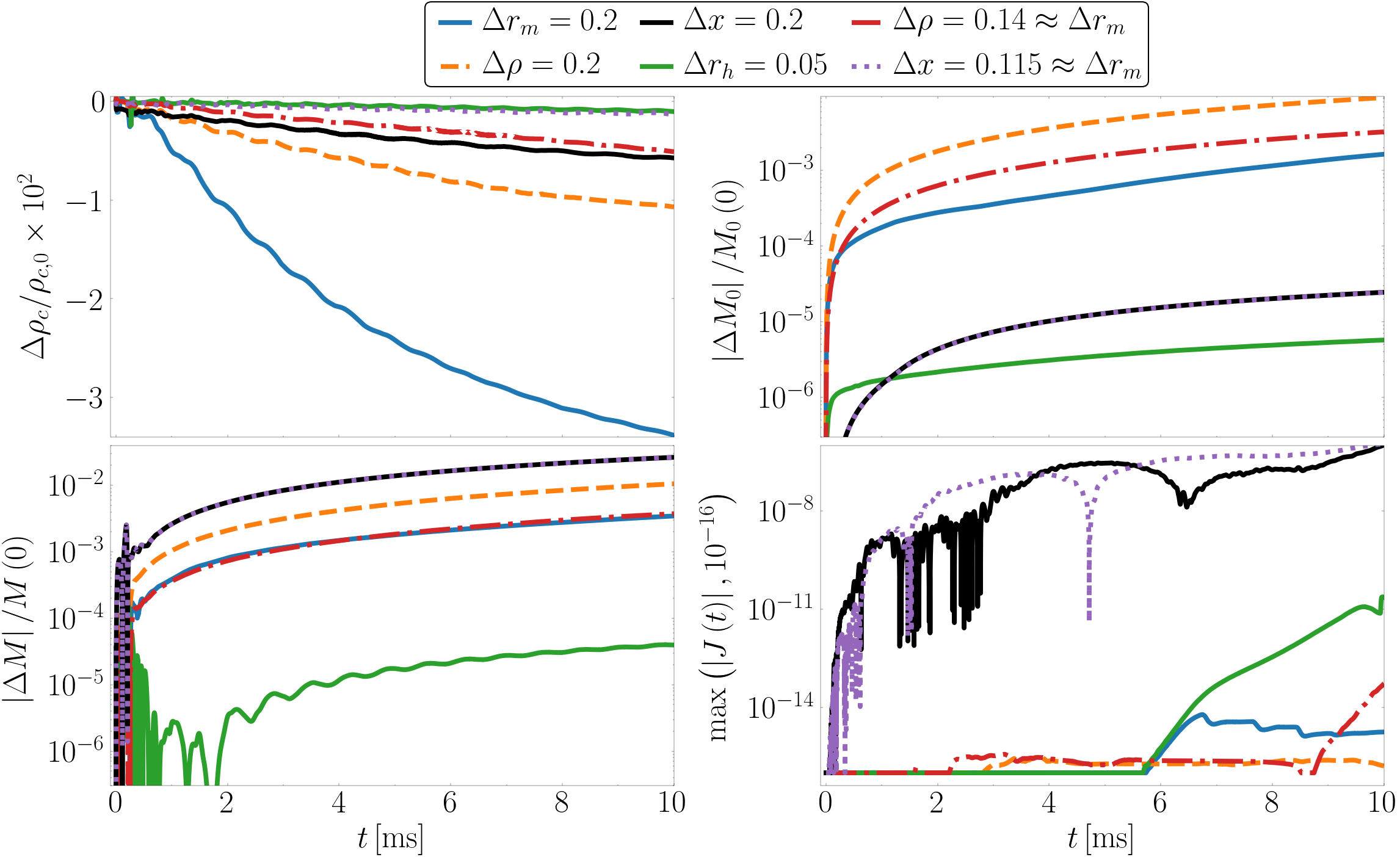}
  \caption{Evolution of a TOV solution simulated in Cartesian, spherical, and cylindrical coordinates. We compare simulations using identical \textit{coordinate} resolutions and identical \textit{effective} resolutions, relative to the reference spherical case ($\Delta r_m = 0.2$). The panels show the time evolution of fractional changes in: (\textbf{top left}) central rest-mass density; (\textbf{top right}) rest mass; (\textbf{bottom left}) gravitational mass; and (\textbf{bottom right}) spin angular momentum. The rest mass, gravitational mass, and spin are computed using \eqref{rest_mass}, \eqref{grav_mass}, and \eqref{spin}, respectively.}
  \label{fig:Dynamical_TOV_compare}
\end{figure*}

We next compare the evolution of identical TOV initial data simulated using our code in spherical and cylindrical coordinates, alongside evolutions in Cartesian coordinates performed with \grhaylhd~\cite{grhayl_github, Cupp_2024}, a pure HD version of \igm based on \grhayl in the \etk. The results are presented in \figref{fig:Dynamical_TOV_compare}. All simulations are performed with the same evolution parameters; only the coordinates and grid resolutions are changed.

To facilitate comparison, we use a spherical coordinates simulation with $\Delta r_m = 0.2$ as our reference case. Simulations in other coordinate systems use either the same resolution (e.g., $\Delta x = 0.2$ for Cartesian coordinates) or the equivalent ``effective'' resolution. For Cartesian coordinates, the effective resolution corresponds to $\Delta x$ such that $\Delta x^2 + \Delta y^2 + \Delta z^2 = 3 \Delta x^2 = \Delta r_m^2$. For the reference $\Delta r_m = 0.2$, this yields an effective $\Delta x \approx 0.11$.

The top left panel of \figref{fig:Dynamical_TOV_compare} shows that the Cartesian simulation at $\Delta x = 0.2$ (solid black) exhibits significantly less diffusion than the spherical simulation at $\Delta r_m = 0.2$ (solid blue). The cylindrical simulation (dashed orange), which combines elements of Cartesian and spherical coordinate systems, falls between the two. When using equivalent effective resolutions, the Cartesian (dotted purple) and cylindrical (dot-dashed red) simulations are even less diffuse, closely approaching the high-resolution spherical results at $\Delta r_h = 0.05$ (solid green).

These results may initially appear counterintuitive, as spherical grids naturally align with the symmetry of the TOV solution. However, the behavior of high-resolution shock-capturing (HRSC) schemes explains this trend. In HRSC schemes, hydrodynamic fluxes are computed across cell faces, and diffusion occurs fastest along coordinate lines. On Cartesian grids, the stellar surface diffuses most rapidly along the $x$, $y$, and $z$ axes, as shown in Fig.~3 of~\cite{WhiskyTHC2}. Diffusion along diagonal lines (e.g., the $x=y=z$ line) is slower, as radial diffusion requires fluxes to traverse multiple orthogonal cell faces in a zigzag pattern. In contrast, on spherical polar grids, radial fluxes align directly with cell faces, enabling faster radial diffusion. This alignment smooths the stellar surface more rapidly, resulting in greater overall diffusion compared to Cartesian grids.

As illustrated in the top left panel of \figref{fig:Dynamical_TOV_compare}, spherical coordinates require a resolution of approximately $\Delta r_h = 0.05$ (solid green) to achieve results comparable to $\Delta x = 0.115$ Cartesian simulations. However, the high-resolution simulation with spherical coordinates requires less than \textit{1/25000} the computational cost, using $400{\times}2{\times}2$ grid points compared to the $346^3$ required for the high-resolution Cartesian simulation.

The top right and bottom panels of \figref{fig:Dynamical_TOV_compare} show the time evolution of the fractional change in rest mass $M_0$, gravitational mass $M$, and spin angular momentum $J$, which we compute using
\begin{align}
    M_0 &= \int d^3x \sqrt{\gamma} \alpha u^0 \rho, \label{rest_mass} \\ 
    M   &= \frac{1}{16 \pi} \int d^3x \sqrt{\bar{\gamma}} \biggl( 16\pi\psi^5 \rho_\epsilon  \nonumber \\
        &+ \psi^{-7} \bar{A}_{ij} \bar{A}^{ij} - \psi\bar{R} - \frac{2}{3} \psi^5K^2\biggr), \label{grav_mass} \\
    J   &= \int d^3x T^0_\phi \alpha \sqrt{\gamma}, \label{spin}
\end{align}
where $\rho_\epsilon = \left( 1 + \epsilon \right)$ is the mass---energy density, $\bar{A}^{ij}$ is the conformal trace-free part of the extrinsic curvature, and $\bar{R} = \gamma^{ij} \bar{R}_{ij}$, with $\bar{R}_{ij}$ being the conformal Ricci tensor. Note that \eqref{spin} is only valid for systems with axisymmetry.

The top right panel of \figref{fig:Dynamical_TOV_compare} shows that the runs in Cartesian coordinates conserve rest-mass to at least the level of truncation error associated with our midpoint-rule approximation for the integral in \eqref{rest_mass}. Although the simulations in cylindrical coordinates and the reference simulation in spherical coordinates do not conserve rest-mass as effectively, the high-resolution simulation in spherical coordinates achieve a comparable level of conservation. As expected, the cylindrical coordinate results are the least accurate due to the system's geometry not conforming well to the star. These inaccuracies are compounded by truncation errors from the source terms in \eqref{eq:basic_evol_expanded}; however, the errors demonstrably converge away with increasing resolution.

In the bottom left panel, the Cartesian runs exhibit the poorest conservation of gravitational mass, as conservation is not guaranteed in this case, while the high-resolution simulation in spherical coordinates again does the best. Finally, the bottom right panel shows that while all simulations exhibit reasonable behavior in the evolution of the spin angular momentum (which should ideally remain at the round-off level), the curvilinear coordinate simulations are significantly closer to this ideal level before outer boundary effects become apparent.

\subsubsection{Non-Rotating NS with a Tabulated \eos}

Here we model a TOV solution with a tabulated \eos in spherical coordinates within the framework of full general relativity---an approach, to the best of our knowledge, undertaken for the first time. We construct initial data using a beta-equilibrium, $T = 0.01\,\mathrm{MeV}$ constant-temperature slice of the SLy4 tabulated \eos~\cite{Schneider_2017}. Atmospheric values for the primitives are set to $\rho_\text{atm} = 1.4 \times 10^{-12}$, $T = 0.01\,\mathrm{MeV}$, and $\ye = 0.5$. For the specific TOV model considered, we select a gravitational mass of $M = 1.4$, corresponding to a central density of $\rho_c = 1.42 \times 10^{-3}$ and a radius of $R = 6.5$.

\begin{figure}[htbp!]
  \centering
  \includegraphics[width=\linewidth]{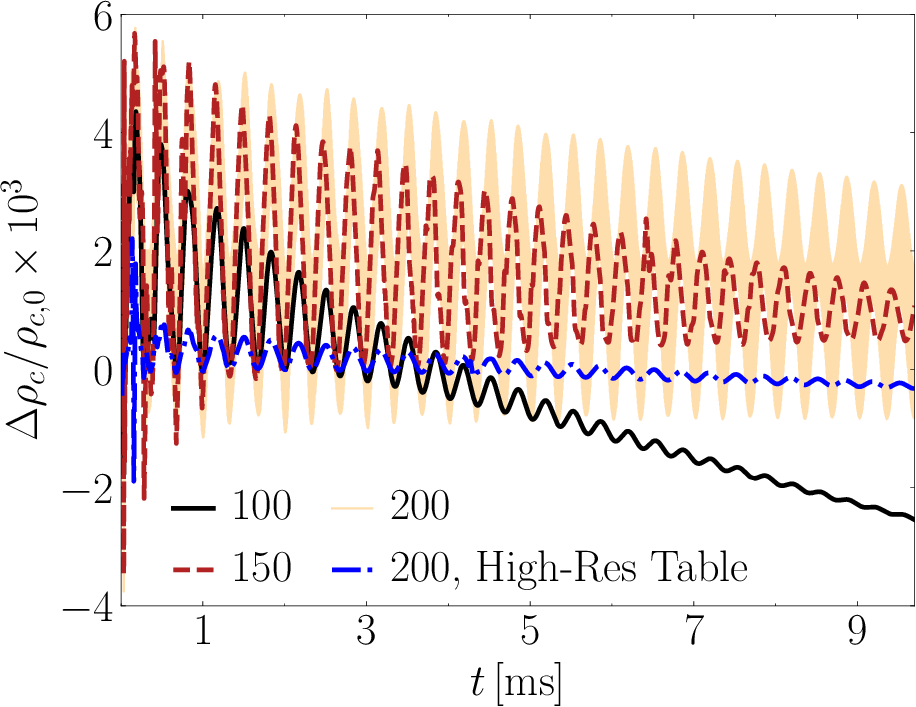}
  \caption{Evolution of the normalized change in central density over time for a TOV model using the SLy4 tabulated \eos in spherical coordinates. We employ radial resolutions of $N_r = \left(100, 150, 200\right)$ while fixing the angular resolution at $N_\theta = N_\phi = 2$. The \eos table resolution is varied to illustrate the emergence and resolution of unphysical high-frequency oscillations.}
  \label{fig:Dynamical_TOV_TabEoS}
\end{figure}

The star is modeled in spherical coordinates with radial resolutions $N_r = (100, 150, 200)$ and angular resolutions set to the minimum allowed in \nrpy: $N_\theta = N_\phi = 2$. For primitive recovery, we again use the one-dimensional routine of~\cite{Newman_2014}. \figref{fig:Dynamical_TOV_TabEoS} presents the results of this test at varying resolutions. Initially, the high-resolution run exhibits problematic high-frequency oscillations in the central density, attributed to insufficient numerical viscosity and the low resolution or inherent non-smoothness of table quantities. However, these oscillations are mitigated when a higher resolution \eos table is used (blue curve), in clear contrast to the oscillatory behavior observed with the lower resolution table (orange curve).

We generated this new table using the SRO code~\footnote{Available at \url{http://stellarcollapse.org}.}, modifying only the sampling options. Specifically, we increased the steps per decade in $\left( \ye, n, T \right)$, where $n$ is the number density, from $\left( 67, 30, 30 \right)$ to $\left( 88, 80, 80 \right)$ in the final table. In future work, we aim to implement a spectral representation for piecewise polytropic and tabulated \eoss~\cite{Lindblom_2010, Lindblom_2022, Foucart_2019, Knight_2023}, which will help address the challenges associated with using raw tables.

\subsubsection{Uniformly Rotating NS with a Hybrid \eos}
\label{uniform_rotation}

In our final test, we examine \groovy's capability to model a uniformly rotating neutron star (NS). This is particularly relevant for potential ``handoff'' studies, where the code could model binary neutron star (BNS) or black hole-neutron star (BHNS) post-merger remnants over seconds-long timescales. For this test, we evolve the BU2 model from Ref.~\cite{Stergioulas_2004} with $K=100$ and $\Gamma=2$. This model describes a rotating NS with the following properties: $M = 1.47$, equatorial coordinate radius $R_e = 8.54$, and dimensionless spin angular momentum $J/M^2 = 3.19 \times 10^{-1}$. Initial data for this model is generated using the open-source \fuka initial data solver~\cite{Papenfort2021a}.

\fuka is based on an extended version of the \kadath spectral solver library, which is specifically designed for solving numerical relativity problems~\cite{Grandclement2009}. \kadath employs a novel non-overlapping multi-grid spectral domain decomposition such that regions of high physical gradients such as the stellar surface can be defined along a domain boundary using appropriate boundary conditions. This method facilitates robust convergence and reduces numerical artifacts such as the \textit{Gibbs phenomenon} produced by modeling shocks and steep gradients with spectral methods. \fuka computes initial data solutions using the eXtended Conformal Thin Sandwich (XCTS) decomposition of Einstein's constraint equations~\cite{York1998,Pfeiffer2002}. Finally, \fuka adopts the conformal flatness approximation ($\gamma_{ij} = \psi^4 \delta_{ij}$ in Cartesian coordinates) and maximal slicing ($K = 0$) conditions. 

For isolated neutron stars, \fuka models matter as an isentropic fluid in equilibrium and co-moving with the fluid undergoing uniform rotation which is parameterized by a rotational velocity, $\omega$ (see, e.g.,~\cite{Papenfort2021a, Tichy2011}). To close the system of equations, an \eos must be specified. For this test, we use a simple polytropic \eos. However, \fuka also supports piecewise polytropic and 1D cold tabulated \eoss for modeling isentropic fluids, which may be adopted in future studies. In this work, we have leveraged \fuka's native Python interface that enables full access to the initial data solution from Python, which we have incorporated into our \nrpy workflow.

PPM's lower-order accuracy at extrema is undesirable when evolving neutron stars in either Cartesian~\cite{Duez_IGM} or spherical coordinates~\cite{Mewes_2020}. Therefore, for this test we use MC reconstruction, which exhibits more consistent---though second-order---convergence with increasing resolution. The spacetime evolution settings remain consistent with those used in the previous subsections. A uniform spherical grid is adopted for this test, with the outer boundary set at $r_\mathrm{max} = 25$. Spatial resolutions of $\left( N_r, N_\theta, N_\phi \right) = \left(300, 12, 2 \right)$, $\left(400, 16, 2 \right)$, and $\left(500, 20, 2 \right)$ are employed, while maintaining a constant density atmosphere of $\rho_{\rm atm} = 1.29 \times 10^{-11}$.

\begin{figure*}[!ht]
  \centering
  \includegraphics[width=1.0\textwidth]{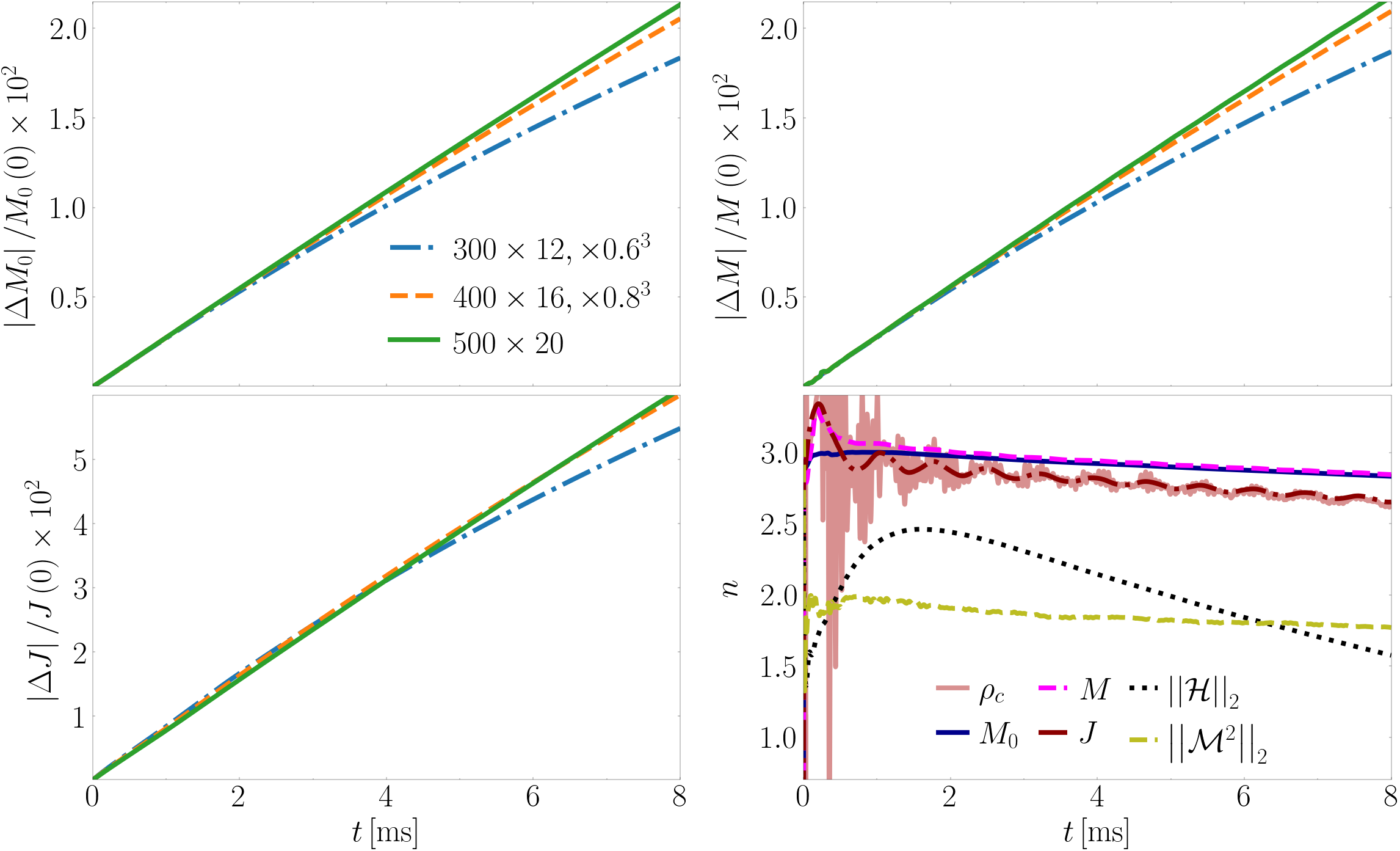}
  \caption{Convergence study for the evolution of a uniformly rotating neutron star using three spatial resolutions: $\left( N_r, N_\theta, N_\phi \right) = \left(300, 12, 2 \right)$, $\left(400, 16, 2 \right)$, and $\left(500, 20, 2 \right)$. \textbf{Top left, top right, and bottom left:} Evolution of the normalized change in rest mass, gravitational mass, and spin angular momentum, respectively. The results from the lower-resolution simulations are rescaled assuming third-order convergence to demonstrate error reduction relative to the high-resolution run. \textbf{Bottom right:} The convergence order $n$, calculated using \eqref{conv_order} with data from the medium and high-resolution simulations. $n$ is plotted for central density, rest and gravitational mass, spin angular momentum, and the $L^2$ norms of the Hamiltonian and momentum constraint violations. The observed convergence orders are consistent with the numerical scheme.}
  \label{fig:Uniform_Rot_NS}
\end{figure*}

For this test we also monitor the $L^2$ norms of the Hamiltonian $\mathcal{H}$ and momentum $\mathcal{M}^2 = \mathcal{M}^i \mathcal{M}_i$ constraint violations, computed using Eqs.\,$\left(46{\text -}47\right)$ from \cite{Ruchlin_2018}. The top left, top right, and bottom left panels of \figref{fig:Uniform_Rot_NS} show the normalized change in rest mass, gravitational mass, and spin angular momentum, respectively. In the high-resolution simulation, the rest and gravitational masses change by no more than 2.5\% over 8 ms, while the angular momentum changes by less than 6\%. Further, we rescale the results from the lower resolution simulations assuming a third-order convergence order. Given that the spacetime evolution employs a fourth-order finite difference scheme, the hydrodynamic finite-volume scheme is approximately second order, while reconstruction schemes cannot be better than first-order at shocks like the NS's surface, and both are coupled to a fourth-order timestepping scheme, the expected convergence order of our numerical scheme should fall between first and fourth order. This can be seen in the bottom right panel of \figref{fig:Uniform_Rot_NS}, where we use data from our medium and high resolution simulations to plot the computed convergence order $n$ of the normalized change in central density, rest- and gravitational mass, spin angular momentum, and the Hamiltonian and momentum constraints. Assuming our scheme is truncation-error dominated, we compute $n$ using
\begin{align}
    n = \frac{1}{\log_{10}\left( f\right)} \log_{10} \left( \frac{\left|\varepsilon \right|^{\mathrm{med}}}{\left|\varepsilon \right|^{\mathrm{high}}}\right),
    \label{conv_order}
\end{align}
where $f$ is the ratio of the grid spacing of the high and medium resolution simulations, and $\varepsilon$ is the computed error at the given resolution. In the continuum limit towards infinite resolution, $\varepsilon$ should converge to zero at some order between one and four. We observe that after initial transient behavior these quantities all converge at expected order. We note that the downward drift in the convergence order of the Hamiltonian constraint is most likely due to our use of a fairly close outer boundary, which is expected considering that the simulations last for over 30 light-crossing times.

\section{Conclusions and Future Work}
\label{conclusion}

In this paper, we introduced \groovy, a numerical relativity code for evolving GRHD fluids with advanced \eoss in diverse coordinate systems, including singular curvilinear (e.g., spherical and cylindrical) and non-singular (Cartesian) coordinates. This development broadens our ability to simulate a wide range of astrophysical systems. Building on the open-source, \igm-based~\cite{Etienne_IGM,Werneck_IGM} \grhayl~\cite{grhayl_github, Cupp_2024}, \groovy dynamically integrates the core GRHD equations and extends them to include key physical processes such as lepton number conservation and neutrino leakage. Its flux-conservative formulation mitigates numerical instabilities near coordinate singularities by solving the equations within an orthonormal basis. The versatility and robustness of \groovy were validated through a comprehensive set of tests in flat, static curved, and dynamical curved spacetimes.

In flat and static spacetime scenarios, \groovy demonstrated its ability to accurately capture shock dynamics in hydrodynamical flows, and model both optically thin and thick gases using a neutrino leakage scheme. For curved, static and dynamical spacetimes, it successfully modeled non-rotating neutron stars with both polytropic and tabulated equations of state, as well as uniformly rotating neutron stars in spherical coordinates. In dynamical spacetime tests, \groovy integrated seamlessly with our BSSN solver in the moving-puncture gauge, showcasing its capability to evolve spacetime metrics coupled with hydrodynamic fields. 
These results establish this preliminary version of \groovy as a first step towards developing a powerful tool for simulating binary neutron star (BNS) and black hole-neutron star (BHNS) post-merger remnants.

In the future, we plan to implement a mean-field prescription for magnetic field evolution~\cite{Most_2023, DelZanna_2022, Radice_2017, Radice_2020, Palenzuela_2021}. This addition will allow us to study the large-scale effects of magnetic field dynamos on the overall dynamics in a primarily hydrodynamic context, as demonstrated in~\cite{Radice_2017}. We also intend to leverage the in-development multi-patch, GPU, and \texttt{Charm++} parallelization infrastructures currently being integrated into \bhah to conduct efficient full 3D simulations of important astrophysical phenomena. 

Ultimately, our goal is to fully incorporate MHD, equipping \groovy with another essential component to accurately capture the critical physics of compact object mergers. By enabling long-term simulations in coordinate systems conforming to the near-symmetries of compact binary remnants, \groovy serves as a powerful and computationally efficient tool for investigating phenomena such as gamma-ray bursts, nucleosynthesis, and the evolution of post-merger remnants. This capability has the potential to drive new discoveries and deepen our understanding of compact astrophysical systems.

\begin{acknowledgments}
\label{acknowledgments}

The authors are grateful to K.~D.~Kokkotas for providing normal mode oscillation frequencies computed using his linearized dynamical-spacetime TOV perturbation code. We also thank T.~W.~Baumgarte and V.~Mewes for fruitful discussions and their valuable insights. TPJ acknowledges support from NASA FINESST-80NSSC23K1437, the Southern Regional Education Board's Dissertation Award program, and West Virginia University's Chancellor's Scholarship program. ZBE's work was supported by NSF grants OAC-2004311, OAC-2411068, AST-2108072, PHY-2110352/2508377, and PHY-2409654, as well as NASA ISFM-80NSSC21K1179, ATP-80NSSC22K1898, and TCAN-80NSSC24K0100. SC and ST thank the University of Idaho P3-R1 Initiative for their support. This research made use of Idaho National Laboratory's High Performance Computing systems located at the Collaborative Computing Center and supported by the Office of Nuclear Energy of the U.S. Department of Energy and the Nuclear Science User Facilities under Contract No. DE-AC07-05ID14517. Computational resources were also partially provided by West Virginia University's Thorny Flat high-performance computing cluster, funded by NSF MRI Award 1726534 and West Virginia University. This work made extensive use of the open-source packages NumPy~\cite{NumPy}, SciPy~\cite{SciPy}, SymPy~\cite{SymPy}, Matplotlib~\cite{Matplotlib}, \grhayl~\cite{grhayl_github}, and \fuka~\cite{fuka_repo}.

\end{acknowledgments}

\appendix

\section{Detailed Analysis of Algorithmic Enhancements in \groovy over \igm}
\label{appendix:igm_groovy_differences}

\subsection{Finite-Differencing Accuracy Study}
\label{appendix:finite_difference_study}

\begin{figure*}[!htbp]
  \centering
  \includegraphics[width=0.9\textwidth]{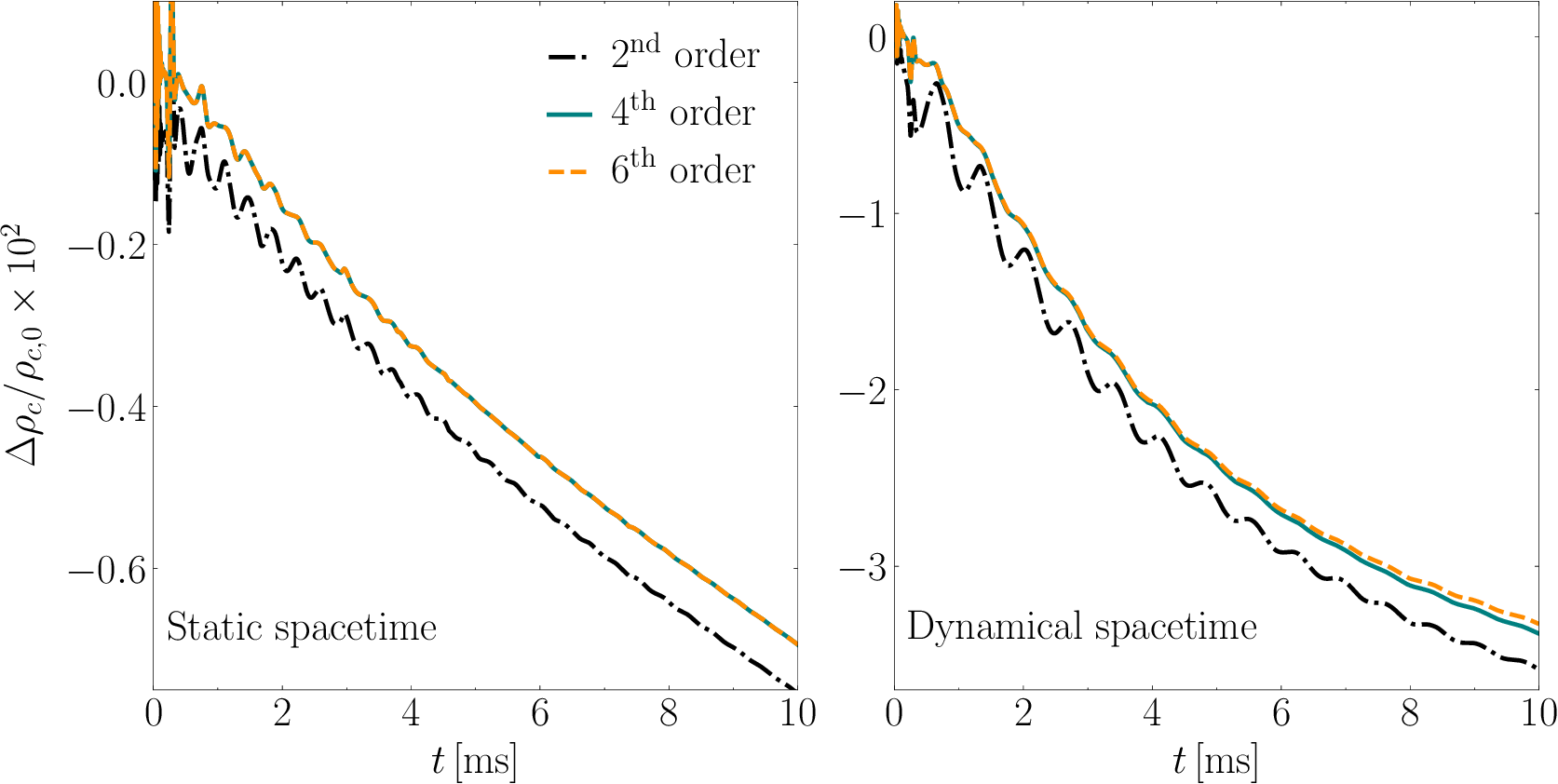}
  \caption{Impact of GRHD metric source term finite differencing order on diffusion of TOV stars. \textbf{Left}: Drift of central density holding the spacetime evolution fixed (Cowling approximation). \textbf{Right}: Same as left panel, but evolving the spacetime.}
  \label{fig:TOV_HO}
\end{figure*}

We evolve the same Tolman-Oppenheimer-Volkoff (TOV) initial data described in Sec.~\ref{code_tests:curved_static} in both static and dynamical spacetimes to assess the impact of second-, fourth-, and sixth-order finite-difference stencils on stellar diffusion in source term calculations. The results, presented in Fig.~\ref{fig:TOV_HO}, show that increasing the finite-differencing order from second to fourth significantly reduces central density drift in both static and dynamical spacetime evolutions. However, further increasing the order to sixth provides negligible improvements, indicating diminishing returns beyond fourth-order accuracy. 

These findings suggest that fourth-order stencils strike a practical balance between computational cost and accuracy for equilibrium neutron star simulations. Our results confirm that the higher-order stencils supported by \nrpy significantly outperform the second-order stencils used in the original \igm implementation, reducing the rate at which the star diffuses.

\subsection{Pressure Floor Study}
\label{appendix:pressure_floor_study}

\begin{figure}[!htbp]
  \centering
  \includegraphics[width=\linewidth]{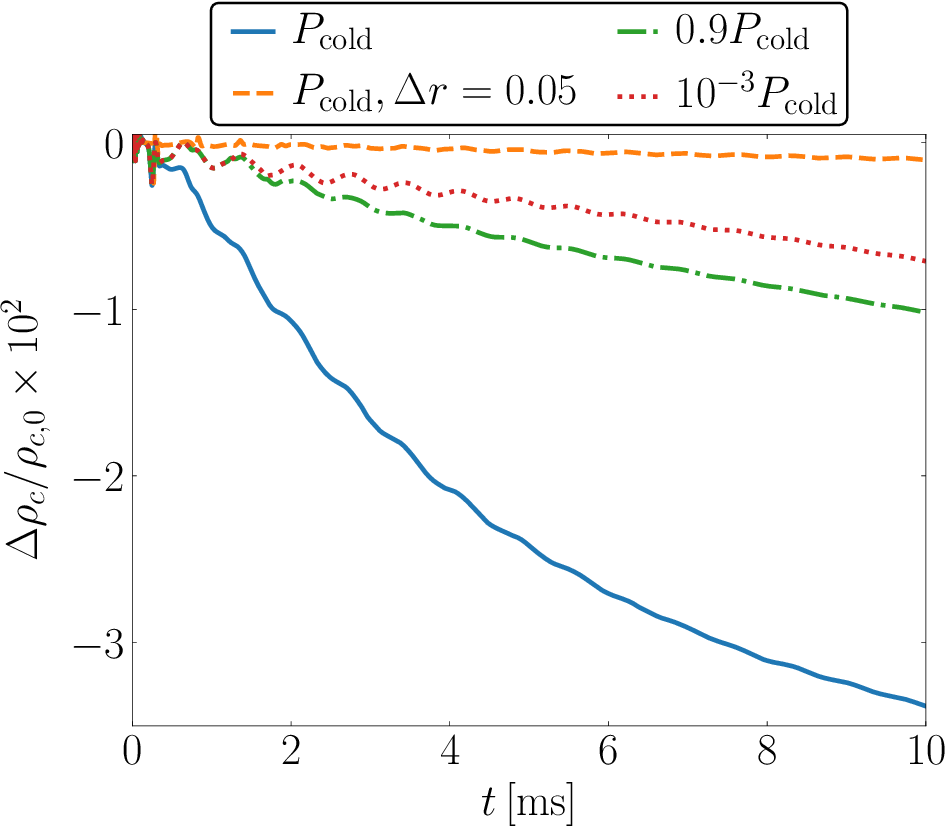}
  \caption{Comparison of the effects of different pressure floors on TOV stellar models using a hybrid \eos, with evolutions performed in spherical coordinates. The blue, green, and red curves represent simulations at a medium coordinate resolution of $\Delta r_m = 0.2$, while the orange curve corresponds to a high-resolution simulation with $\Delta r_h = 0.05$.}
  \label{fig:Dynamical_TOV_cold}
\end{figure}

\figref{fig:Dynamical_TOV_cold} illustrates the impact of varying pressure floors on the evolution of polytropic TOV initial data in spherical coordinates, using the same numerical setup as in \secref{code_tests:dynamical_spacetime_hybrideos_tov}. The blue, green, and red curves correspond to simulations at a medium resolution of $\Delta r = 0.2$, while the orange curve represents a high-resolution run at $\Delta r = 0.05$. For the hybrid \eos used here, the correct pressure floor is the cold pressure, $P_{\rm cold}$, defined by the $\Gamma=2, K=100$ polytrope.

While setting the floor below $P_{\rm cold}$ introduces artificial cooling, which is not physically consistent with our GRHD equations or \eos, it reduces stellar diffusion, as shown in the figure. Despite these numerical benefits, lowering the floor sacrifices physical realism. Therefore, we choose $P_{\rm cold}$ as the pressure floor in our simulations to maintain physical fidelity.

\bibliographystyle{apsrev4-1}
\bibliography{references}

\end{document}